%% file: Final_Version.tex
\documentclass[10pt, final, journal, letterpaper, twocolumn]{IEEEtran}
\IEEEoverridecommandlockouts
\usepackage[T1]{fontenc}
\newif\ifhavebib

\havebibtrue
\usepackage{latexsym}
\usepackage{setspace}
\usepackage{bm}
\usepackage{lipsum}
\usepackage{bbm}
\input{begin.tex}
\begin{document}
\providecommand{\keywords}[1]{\textbf{\textit{Index terms---}} #1}

\title{Blind Graph Matching Using Graph Signals}

\IEEEoverridecommandlockouts
\author{
	Hang Liu,~\IEEEmembership{Member,~IEEE}, Anna~Scaglione,~\IEEEmembership{Fellow,~IEEE}, and
	Hoi-To~Wai,~\IEEEmembership{Member,~IEEE}
	\thanks{This work was supported in part by the DoD-ARO under Grant No. W911NF2010153 and Grant No. W911NF2210228. Part of this work was presented at the  IEEE International Workshop on Computational Advances in Multi-Sensor Adaptive Processing (CAMSAP), Costa Rica, December 2023 \cite{10403434}. 
 
 H. Liu and A. Scaglione are with the 
		Department of Electrical and Computer Engineering, Cornell Tech, Cornell University, New York, NY, 10044 USA (e-mail: \{hl2382, as337\}@cornell.edu).
		H.-T. Wai is with the Department of SEEM, The Chinese University of Hong Kong, Hong Kong. (e-mail: htwai@se.cuhk.edu.hk).
	}
}

\maketitle
\begin{abstract}
Classical graph matching aims to find a node correspondence between two unlabeled graphs of known topologies. This problem has a wide range of applications, from matching identities in social networks to identifying similar biological network functions across species. However, 
when the underlying graphs are unknown, the use of conventional graph matching methods requires  inferring the graph topologies first, a process that is highly sensitive to observation errors. In this paper, we tackle the \emph{blind graph matching} problem with unknown underlying graphs directly using observations of graph signals, which are generated from graph filters applied to graph signal excitations. We propose to construct sample covariance matrices from the observed signals and match the nodes based on the selected sample eigenvectors. Our analysis shows that the blind matching outcome converges to the result obtained with known graph topologies when the signal sampling size is large and the signal noise is small. Numerical results showcase the performance improvement of the proposed algorithm compared to matching two estimated underlying graphs learned from the graph signals. 
\end{abstract}

%\IEEEpeerreviewmaketitle
\section{Introduction}
Graph matching refers to the process of finding the node correspondence between two graphs. This problem has attracted widespread attention owing to its vital applications in many fields, such as pattern recognition \cite{GM_PR}, network analysis \cite{7536145}, and computational biology \cite{singh2008global}. Graph matching can be categorized into three main approaches \cite{GMP_Survey}: graph edit distance, graph kernels, and graph embedding.  The most popular of graph embedding methods is spectral embedding, also known as \emph{spectral graph matching}. 

Since our method relates to spectral graph matching, our review of the state of the art will focus on this class.
%we are going to focus on this class in reviewing the state of the art. 
Specifically, spectral graph matching finds proper representations of graphs in the eigenspaces of adjacency or Laplacian matrices, simplifying the original NP-hard combinatorial search problem into a more tractable form \cite{GMP_eign1}. The author in \cite{GMP_eign1} formulated the problem of exact graph matching as finding a permutation between adjacency matrices. It is shown that the optimal permutation can be obtained by first computing the eigendecomposition of adjacency matrices and then solving a bipartite maximum weighted matching problem.  The work in \cite{GMP_eign3} further extended the method in \cite{GMP_eign1} to handle inexact matching of two graphs with different sizes by choosing the top eigenvalues as the projection space. Another extension of \cite{GMP_eign1} is presented in \cite{GMP_eign2}, which considered the eigendecomposition of Laplacian matrices and used eigenvector histograms for alignment. The framework in \cite{GMP_eign2} was further extended in \cite{GMP_eigen5} introducing a local node similarity measure; in the paper, the spectral information on Laplacian matrices is referred to as the global node similarity.  Moreover, \cite{GMP_eigen6} proposed a multi-resolution spectral method. More recently, \cite{GMP_eigen7} proposed a pairwise eigenvector alignment method that was reported to be robust to sign ambiguity and eigenvalue multiplicity. 

%In spectral methods aligning the two eigenbases requires the strict ordering of eigenvalues. Although adjacency matrices of large  Erd\"{o}s-R\'{e}nyi (ER) random graphs tend to have distinct eigenvalues with high probability \cite{GMP_spectrum}, the assumption does not apply in general, as geometric multiplicities are possible.  There are methods in the literature that propose effective heuristics, adding perturbations to repeated eigenvalues \cite{GMP_eign1} and selecting a subset of eigenvectors \cite{GMP_eign2} or the pairwise eigenvector alignment method proposed in , 
% {(to) it seems that our approach cannot handle this issue of repeated eigenvalue as well - since distinct eigenvalue for the graph filter $h(\lambda_i)$ implies distinct $\lambda_i$ - right?} {\color{red} Yes. Please see Section III-A for the update.}

%Except for the above eigendecomposition-based methods, 
Another line of work searches the matching permutation matrix by continuous, often convex, relaxations. Popular choices of the relaxations include relaxing the feasible set to the set of doubly stochastic matrices \cite{GMP_relax,GMP_relax2}, quadratic matrices \cite{DS++} or orthogonal matrices \cite{GMP_eign1}, a non-negative simplex \cite{GMP_relax4}, or the set with a constant Frobenius norm \cite{GMP_relax3}. Moreover, a convex relaxation method for multi-graph matching was studied in  \cite{GMP_relax5}, while \cite{GMP_relax6} proposed a distributed algorithm for graph matching with convex relaxations. 

%All the aforementioned graph matching methods  \cite{GSP1}

Besides designing computationally efficient algorithms for graph matching, another critical problem is determining when finding a good matching is possible at all. The authors in \cite{GMP_relax} studied correlated random Bernoulli graphs and found that the convex relaxation method works only if the correlation between two graphs is sufficiently large. Similarly, \cite{GMP_theory3,GMP_theory2,GMP_theory4,GMP_theory1} studied the condition of successful recovery from an information-theoretical perspective and proved the existence of a sharp phase transition in the recovery of the correct permutations for Gaussian models and Erd\"{o}s-R\'{e}nyi (ER) random graphs. An algorithm that approaches the transiting threshold has been proposed in \cite{GMP_theory5}.
For a more general setup, it has been recognized that graphs with symmetrical structures, such as cycles, do not have a unique matching \cite{GMP_relax2,GMP_sym2}. If symmetries exist, more than one permutation leads to an equally good matching; thus, the optimal one is difficult to identify. 
Also, identifying symmetries in a given graph is challenging. Sufficient conditions to guarantee the graph asymmetry were proposed in \cite{symmetry02,10.1093/imaiai/iav002,GMP_relax2}.
Numerical experiments in \cite{GMP_sym2} reported that large ER random graphs have a very high probability of being asymmetric.  Additionally,  \cite{GMP_sym2} identified that symmetric graphs have two or more subgraphs with the same inner structure and outer connections.

The current work on graph matching assumes prior knowledge of the graph topology. However, in many applications, such as social networks, infrastructure networks, and functional brain connectivity, direct observations of network links are not available. Instead, the underlying graph is constructed from observations of interactions between nodes, known as graph signals. These signals can be opinions in social networks, nodal measurements in infrastructure networks,  encephalography signals in brain connectivity, and gene network expressions due to genetic interactions. When only graph signals are available, a common heuristic for graph matching is first inferring the graph topology from the observed signals by topology inference (a.k.a. graph learning), and then matching nodes based on the estimated topology. However, this heuristic  is prone to errors because topology inference usually requires strong assumptions about graph structures or signals \cite{Topology_inference}.  
On the other hand, recent research has shown that graph analysis can be efficiently carried out using filtered graph signals generated from graph filters \cite{LPGP_mag}. For example, \cite{GSP1,9757839} used filtered graph signals to detect communities and central nodes of unknown graphs.

% However, identifying symmetries with observations of graph signals   remains an open problem.
%\cite{GMP_sym1}

\subsection{Contributions}
In this work, we propose and analyze a blind graph matching method using graph signals, which does not require direct topology inference or prior knowledge of the adjacency or Laplacian matrices. We assume that the two sets of graph signals are generated over non-identical graph filters that exhibit the same low-pass or high-pass graph spectrum trends, which, in turn, means that the graph frequency order is preserved.  Under this relatively mild assumption, we compute the eigenbases of sample covariance matrices from the graph signals and match nodes by finding the correspondence in the eigenbases. Our method can be seen as an extension of the spectral method in \cite{GMP_eign1} to the blind scenario. The contributions of this work are summarized as follows.
\begin{itemize}
	\item We propose a spectral method for matching two unknown graphs using their filtered graph signals. Our approach involves computing the eigenbases of sampling covariance matrices on the two signal sequences and constructing a node similarity measuring matrix based on these eigenbases. We then convert the blind matching task to a linear assignment problem and solve it by the Hungarian method \cite{Hungarian} and the greedy method \cite{1544893}.
	%\item 
%	We propose a sufficient and necessary condition for the uniqueness of optimal graph matching, which can be used to check the identifiability of any graph matching problem.  
%We propose an approximate identifiability check approach for blind graph matching, which can efficiently detect symmetric structures of underlying graphs from graph signals. Our method relies on empirical eigenvectors of the sample covariance matrices and thus does not require knowledge of the graph topology.
%	 is  computationally efficient compared to the exhaustive search for symmetric structures in graphs. 
	\item We analyze the performance degradation in blind graph matching compared to the case where the graph topology is known. Specifically, we quantify the optimality gap in the matching objective and the matching error probability	by analyzing the perturbation to the node similarity matrix caused by signal sampling. Our results show that blind matching achieves diminishing matching error with sufficiently many signal observations and small signal noise, particularly when the number of samples scales proportionally to $n\log n$, where $n$ denotes the graph size.

 \item 
 %Based on the analytical result, we propose a heuristic approach for selecting eigensubspaces in our matching algorithm.
 Our analysis suggests that the precision of blind graph matching is significantly influenced by the spectral gap of the signal covariance matrices. Therefore, selecting a subset of sample eigenvectors can effectively mitigate the impact of perturbations in signal sampling. We propose a heuristic method for eigenvector selection, which enhances matching accuracy while considerably reducing computational time.
\end{itemize}
We conduct simulations on both synthetic data and real-world datasets to verify the efficiency of the proposed method. The results demonstrate that our method is more robust against errors and achieves more accurate matching compared to the heuristic combination of graph topology inference and graph matching.

\subsection{Organization and Notations}
The paper is organized as follows. In Section \ref{sec2}, we introduce conventional spectral graph matching with known graph topologies. In Section \ref{sec3}, we describe the blind graph matching problem and propose our solution to it.  In  Section \ref{sec4}, we analyze the performance of the proposed algorithm and discuss the eigenvector selection scheme.
In Section \ref{sec5}, we present numerical results to evaluate the proposed method. Finally, this paper concludes in Section \ref{sec6}.

Throughout, 
%we use $\Real$ and $\Complex$ to denote the real and complex number sets, respectively. 
we use regular letters, bold small letters, and bold capital letters to denote scalars, vectors, and matrices, respectively. We use  $\Xv^T$ to denote the transpose of matrix $\Xv$, $\overline \Xv$ to denote the matrix containing the absolute value of the entries of $\Xv$,  $\tr(\Xv)$ to denote the trace of $\Xv$, and $\rank(\Xv)$ to denote the rank of $\Xv$.
We use $x_i$ to denote the $i$-th entry of vector $\xv$, $x_{ij}$ or $[\Xv]_{ij}$ interchangeably to denote the $(i,j)$-th entry of matrix $\Xv$,
% $[\Xv]_{i,:}$ to denote the $i$-th row, and $[\Xv]_{:,j}$ or 
and $\xv_j$ to denote the $j$-th column of $\Xv$. 
The real normal distribution with mean $\muv$ and covariance $\Cv$ is denoted by $\Norm(\muv,\Cv)$, and the cardinality of set $\mathcal{S}$ is denoted by $\abs{\mathcal{S}}$.
We use 
$\norm{\cdot}_p$ to denote the $\ell_p$ norm,  $\norm{\cdot}_F$ (resp. $\norm{\cdot}_2$) to denote the matrix Frobenius (resp. spectral) norm, $\Iv_n$ to denote the $n\times n$ identity matrix,  $\bf 1$ to denote the all-one vector with an appropriate size,
and $\diag(\xv)$ to denote a diagonal matrix with the diagonal entries specified by $\xv$. For any positive integer $n$, we denote the factorial of $n$ by $n!$ and define $[n]\triangleq\{1,2,\cdots,n\}$.

\section{Conventional Spectral Graph Matching}\label{sec2}
Consider  two undirected graphs $\mathcal{G}_1=(\mathcal{V}_1,\mathcal{E}_1)$ and $\mathcal{G}_2=(\mathcal{V}_2,\mathcal{E}_2)$, where $\mathcal{V}_i$ and $\mathcal{E}_i$ denote the sets of nodes and edges of the $i$-th graph, $i=1,2$, respectively.  We assume that both graphs have the same number of nodes denoted by $n$.\footnote{The graph matching framework presented in this work can be readily extended to matching two graphs with unequal numbers of nodes by creating dummy nodes at one graph.} Each graph $\mathcal{G}_i, i=1,2$, is associated with a symmetric adjacency matrix $\Av^{(i)}\in\Real_{+}^{n\times n}$, where $a_{kl}^{(i)}=a_{lk}^{(i)}>0$ if and only if $(k,l)\in\mathcal{E}_i$. Note that the model of $\Av^{(i)}$ is applicable to both weighted and unweighted graphs. The Laplacian matrix of Graph $\mathcal{G}_i$ is defined as $\Lv^{(i)}\triangleq \diag(\Av^{(i)}{\bf 1})-\Av^{(i)}$.

The objective of graph matching is to find a mapping between the two node sets $\mathcal{V}_1$ and $\mathcal{V}_2$, a.k.a. graph isomorphism, such that the adjacency relationship is maximally preserved. To achieve this, we search for a bijective node permutation function $\pi(\cdot):[n]\to[n]$ that maps each node $v\in\mathcal{V}_1$ to $\pi(v)\in\mathcal{V}_2$. Denote by $\mathcal{P}_n$ the set of $n\times n$ permutation matrices.
We represent any node permutation $\pi(\cdot)$ as a corresponding permutation matrix by $\Pv\in \mathcal{P}_n$ such that $p_{kl}=1$ if $\pi(k)=l$ and $p_{kl}=0$ otherwise. 
Throughout the paper, we use $\pi(\cdot)$ and $\Pv$ interchangeably to denote the node permutation.
After permuting the nodes of $\mathcal{G}_1$ by any $\Pv$, its Laplacian matrix  can be represented as $\Pv^T\Lv^{(1)}\Pv$. 
Accordingly, the accuracy of graph matching with respect to (w.r.t.) any $\Pv\in \mathcal{P}_n$ can be measured by the following disagreement function \cite{GMP_relax2,GMP_sym2}:\footnote{Note that similar disagreement functions are also used in literature with the Laplacian matrices replaced by adjacency matrices \cite{GMP_eign1} or their normalized versions \cite{GMP_eign2}.} %Our proposed method can be readily extended to problem formulations using these matrices.}
\begin{align}\label{eq1}
	\text{dis}_{\mathcal{G}_1\to\mathcal{G}_2}(\Pv)\triangleq \norm{\Lv^{(2)}-\Pv^T\Lv^{(1)}\Pv}^2_F.
\end{align}  
Note that the measurement in \eqref{eq1} unifies the exact and inexact graph matching problems. In particular, $\mathcal{G}_1$ and $\mathcal{G}_2$ are isomorphic if and only if $	\text{dis}_{\mathcal{G}_1\to\mathcal{G}_2}(\Pv)=0$ for some $\Pv\in\mathcal{P}_n$. 
Motivated by this, graph matching finds the optimal permutation $\Pv^\star$ by minimizing \eqref{eq1} as
\begin{align}\label{eq02}
	\Pv^\star=\argmin_{\Pv\in\mathcal{P}_n} \text{dis}_{\mathcal{G}_1\to\mathcal{G}_2}(\Pv).
\end{align}

\subsection{Spectral Graph Matching}\label{sec_2c}
%After verifying the identifiability condition by Lemma \ref{lemma1}, solving \eqref{eq02} leads to a unique node permutation. However, 
Problem \eqref{eq02} is combinatorial and difficult to solve directly. In this section, we review an approximate solution to \eqref{eq02} known as \emph{spectral graph matching} \cite{GMP_eign1}, which is the basis for the blind graph matching algorithm we propose and study in this paper. 
Let the eigendecomposition of $\Lv^{(i)},i=1,2,$ be
\begin{align}\label{eq03}
	\Lv^{(i)}=\Vv^{(i)}\Gammav^{(i)}(\Vv^{(i)})^T,
\end{align}
where $\Gammav^{(i)}$ is the diagonal matrix with diagonal elements aligning the eigenvalues in descending order $\gamma_1^{(i)}\geq \gamma_2^{(i)}\geq\cdots\geq \gamma_n^{(i)}=0$, and $\Vv^{(i)}\in\Real^{n\times n}$ is the orthogonal matrix containing the corresponding eigenvectors.  

We make the assumption, as done in \cite{GMP_eign1}, that the eigenvalues of each graph, \ie $\{\gamma_k^{(i)}\}_{k=1}^n$, are distinct, which is a prerequisite for the spectral graph matching method to work.
Consider the case of exact matching with  $\mathcal{G}_1$ and $\mathcal{G}_2$ isomorphic, \ie $\text{dis}_{\mathcal{G}_1\to\mathcal{G}_2}(\Pv^\star)=0$ for some $\Pv^\star\in\mathcal{P}_n$. The spectral method first relaxes the feasible set to the set of orthogonal matrices. By substituting \eqref{eq03} into \eqref{eq1}, the optimal orthogonal matrix has the following expression:
\begin{align}\label{eq04}
\Vv^{(1)}\Sv(\Vv^{(2)})^T,
\end{align}
where $\Sv$ is some unknown diagonal matrix with diagonal elements being either $1$ or $-1$. In \eqref{eq04}, $\Sv$ represents the sign ambiguity in the eigendecomposition. 

Due to the combinatorial nature of $\Sv$, it is difficult to directly compute the permutation matrix in \eqref{eq04}. Denote by $\overline \Vv^{(i)}$ the matrix containing the absolute value of the entries of $\Vv^{(i)}$, \ie $ [\overline \Vv^{(i)}]_{lk}=|v_{lk}|,\forall l,k$. Applying the triangle inequality, for $\forall \Pv\in\mathcal{P}_n$: 
\begin{align}\label{eq5a}
	\tr(	\Pv^T\Vv^{(1)}\Sv(\Vv^{(2)})^T)\leq \tr(	\Pv^T\overline\Vv^{(1)}(\overline\Vv^{(2)})^T),
\end{align}
where the equality holds if the graphs are isomorphic and $\Pv=\Pv^\star$. 
Furthermore, we bound the right-hand side (r.h.s.) of \eqref{eq5a} as 
\begin{align}\label{eq5b}
	\tr(\Pv^T\overline\Vv^{(1)}(\overline\Vv^{(2)})^T)
%	=&\sum_{j=1}^n\sum_{k=1}^K |v_{\pi(j)k}^{(1)}||v_{jk}^{(2)}|\nonumber\\
	=&\sum_{j=1}^n (\overline\vv_j^{(1)})^T(\Pv\overline\vv_j^{(2)})\nonumber\\
	\overset{(a)}{\leq}&\sum_{j=1}^n \norm{\overline\vv_j^{(1)}}_2\norm{\Pv\overline\vv_j^{(2)}}_2=n,
\end{align}
where $(a)$ follows from the Cauchy–Schwarz inequality, and the equality in $(a)$ holds if $\Pv=\Pv^\star$. 
Leveraging \eqref{eq5a} and \eqref{eq5b}, it is expected that optimizing the r.h.s. of \eqref{eq5a} provides a promising solution to \eqref{eq02}, as the maximum on both sides of \eqref{eq5b} is attained at $\Pv=\Pv^\star$ \cite{GMP_eign1}. Motivated by this, \cite{GMP_eign1} proposed to compute the permutation matching matrix as:
\begin{align}\label{eq05}
	\Pv^{\star\star}=\argmax_{\Pv\in\mathcal{P}_n} \tr(	\Pv^T\overline\Vv^{(1)}(\overline\Vv^{(2)})^T).
\end{align}

The solution $\Pv^{\star\star}$ in \eqref{eq05} is optimal to \eqref{eq02} when the two graphs are exactly isomorphic. Otherwise, we have $\Pv^{\star\star} \approx\Pv^\star$ for inexact matching with two nearly isomorphic graphs \cite{GMP_eign1}. 
Problem \eqref{eq05} is a linear assignment problem and can be efficiently solved by existing solvers, such as the Hungarian method \cite{Hungarian}.

As a final remark, we note that the unique ordering of the eigenvectors in $\Vv^{(i)}$ plays a critical role in spectral graph matching.  Specifically, the formulation in \eqref{eq05} requires the eigendecompositions of $\Lv^{(1)}$ and $\Lv^{(2)}$ have the same order of the eigenvalues. This condition is fulfilled with distinct eigenvalues in the decomposition. However, as we shall demonstrate in the subsequent section, having a unique and identical ordering of eigenvalues  is essential for blind graph matching, but this cannot always be guaranteed with unknown Laplacian matrices.

\section{Blind Graph Matching}\label{sec3}
\subsection{System Model}\label{sec_3a}
We assume that neither the graph topology nor the information on adjacency/Laplacian matrices is available. Instead, we observe two sequences of signals $\{\yv^{(i)}_m\}_{m=1}^M$ over the two graphs known as \emph{filtered graph signals}. The filtered graph signals of each graph $\mathcal{G}_i$ are generated by a graph filter, which is a matrix polynomial of the Laplacian matrix $\Lv^{(i)}$ as\footnote{In this work, the graph filter is defined as a function of the Laplacian matrix, also known as the graph shift operator (GSO). We emphasize that our method is flexible and can be extended to other forms of GSOs. For example, if the graph filter is formulated as a polynomial of the adjacency matrix, our proposed method can be adapted by replacing $\Lv^{(i)}$ with $\Av^{(i)}$.}
\begin{align}\label{eq06}
	\mathcal{H}_i(\Lv^{(i)})=\sum_{t=0}^{T_d-1} h_t^{(i)} (\Lv^{(i)})^t=\Vv^{(i)}\left( \sum_{t=0}^{T_d-1}h_t^{(i)}(\Gammav^{(i)})^t\right) (\Vv^{(i)})^T,
\end{align}
where $T_d$ is the order of the graph filter, and $\{h_t^{(i)}\}$ are the filter coefficients. 
With \eqref{eq06}, the observed signal vector $\yv^{(i)}_m\in\Real^{n\times 1}$ is the output of the graph filter subject to certain excitation signals $\xv^{(i)}_m\in\Real^{n\times 1}$, as
\begin{align}\label{eq_signal}
	\yv^{(i)}_m=	\mathcal{H}_i(\Lv^{(i)})	\xv^{(i)}_m+\wv^{(i)}_{m}, i=1,2, m=1,\cdots,M,
\end{align}
where $\wv^{(i)}_{m}$ represents the modeling error and measurement noise following the distribution of $\Norm({\bf 0},\sigma^2\Iv_n)$. We assume that $\xv_m$ satisfies $\E[\xv_m^{(i)}]=\bf 0$ and  $\E[\xv^{(i)}_m(\xv^{(i)}_m)^T]=\Iv_n,\forall m$.

If both graphs employ an identical graph filter, \ie $	\mathcal{H}_1(\cdot)=\mathcal{H}_2(\cdot)$, we can directly extend \eqref{eq02} to the blind graph matching scenario by replacing the true Laplacian matrices with the sample covariance matrices of the filtered graph signals. However, since a graph filter characterizes how local graph structures affect the corresponding signal models, the filters $	\mathcal{H}_1$ and $	\mathcal{H}_2$ for two different graphs are generally non-identical.  
In this work, we assume that $	\mathcal{H}_1$ and $	\mathcal{H}_2$ are similar in the sense that \emph{they preserve the same unique ordering in the graph spectral domain}. Specifically, 
we see from \eqref{eq06} that the eigenvalues of  $	\mathcal{H}_i(\Lv^{(i)})$, a.k.a. the frequency responses, are given by 
\begin{align}\label{eq09}
	\tilde h_k^{(i)}=\sum_{t=0}^{T_d-1}h_t^{(i)}(\gamma^{(i)}_k)^t,1\leq k\leq n.
\end{align}
Accordingly, for each $i=1,2,$ we sort the absolute value of $\tilde h_k^{(i)}$  in descending order to obtain an associated index ordering function. 
%$o^{(i)}(\cdot):[n]\to [n]$. In other words, $\tilde h_k^{(i)}$ has the $o^{(i)}(k)$-th largest magnitude, and for any $ k,k^\prime\in [n]$ satisfying $o^{(i)}(k)> o^{(i)}(k^\prime)$, we have $|\tilde h_k^{(i)}|< |\tilde h_{k^\prime}^{(i)}|$.  
The assumption is summarized as follows.
\assumption{\label{ass2} The ordering functions of the two sets of frequency responses are the same.
%\ie $o^{(1)}(k)=o^{(2)}(k)= o(k),\forall 1\leq k\leq n$. 
Moreover, the frequency responses of each graph filter have distinct magnitudes, \ie $|\tilde h_k^{(1)}|\neq |\tilde h_{k^\prime}^{(1)}|$ and $|\tilde h_k^{(2)}|\neq |\tilde h_{k^\prime}^{(2)}|, \forall k\neq k^\prime$.
}

Assumption \ref{ass2} can be satisfied, for instance, when the two graph filters exhibit the same low-pass or high-pass tendency. Examples of such filters  are provided below.
\example[Low-pass graph filter]{\label{ex1}Low-pass graph filters concentrate their frequency responses at low graph frequencies. Examples include $\mathcal{H}(\Lv)=(\Iv_n-\a\Lv)^{T_d}$ and $\mathcal{H}(\Lv)=(\Iv_n+\a\Lv)^{-1}$ with $T_d>0$ and $\a>0$, which are widely adopted in diffusion processes and dynamic models  \cite{LPGP_mag}. 
}
\example[High-pass graph filter]{The auto-regressive moving average filter that is frequently used in graph neural networks $\mathcal{H}(\Lv)=\a_1(\Iv_n-\a_2(\Iv_n-\Lv))^{-1}$ is high-pass with  $\a_1>0$ and $\a_2<0$  \cite{9336270}.
}
\subsection{Blind Graph Matching}
\begin{figure}[!t]
	\centering
	\includegraphics[width=3.4 in]{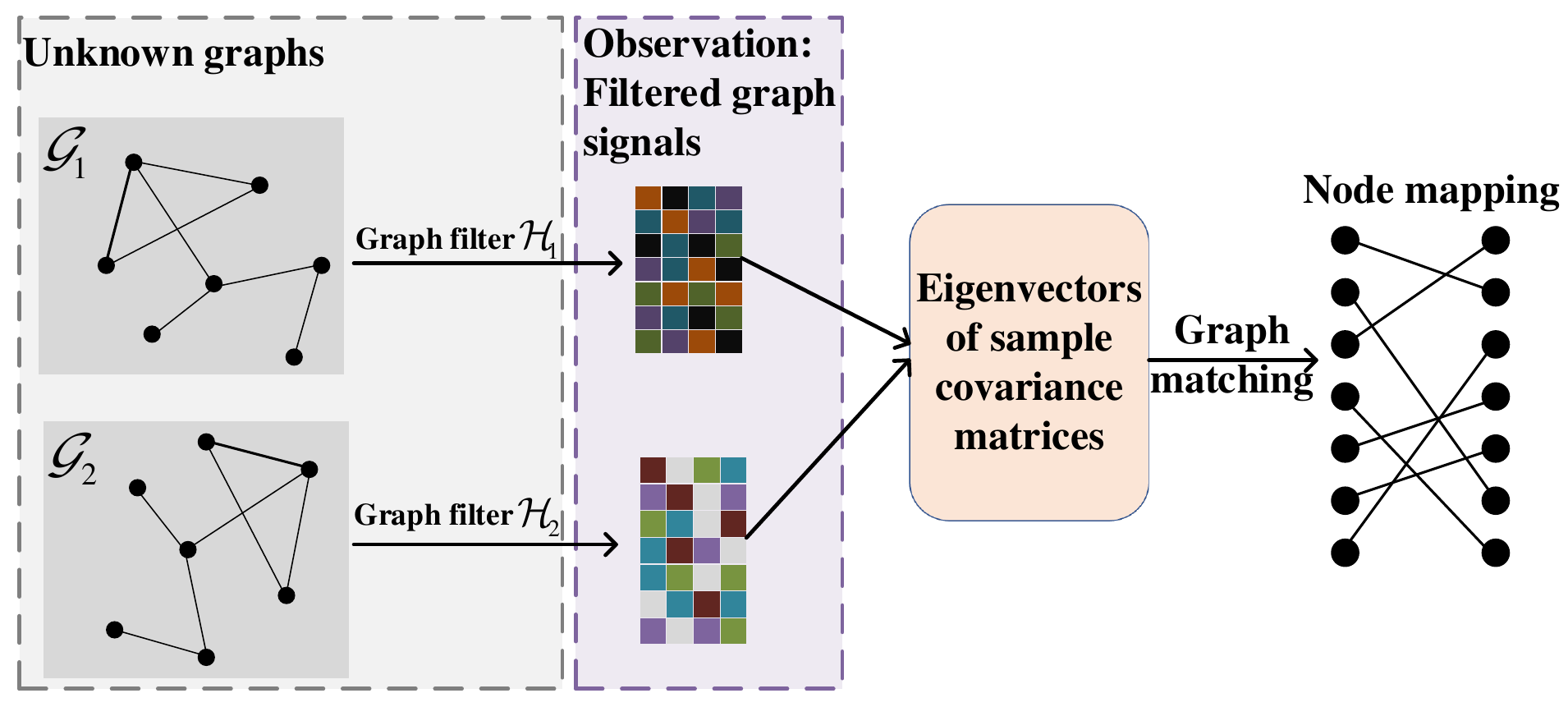}
	\caption{Overview of the blind graph matching approach.}
	\label{fig_sys}
\end{figure}
As depicted in Fig. \ref{fig_sys}, we compute the sample covariance matrix of the $M$ filtered graph signals $\{\yv^{(i)}_m\}_{m=1}^M$ by
\begin{align}\label{eq10}
	\widehat \Cv^{(i)}_y=\frac{1}{M} \sum_{m=1}^M\yv^{(i)}_m(\yv^{(i)}_m)^T-\sigma^2\Iv_n, i=1,2.
\end{align}
In \eqref{eq10}, we assume that  $\sigma^2$ is known, utilizing it to adjust for bias in the sample covariance estimation. In situations where  $\sigma^2$ is unknown, it can be estimated through statistical methods, such as spectrum filtering and median absolute deviation. Denote the noiseless covariance matrix of $\yv^{(i)}_m$ by $	\Cv^{(i)}_y$. The bias correction technique in \eqref{eq10} ensures that $\E[\widehat \Cv_y^{(i)}]= \Cv_y^{(i)}$. From \eqref{eq06}--\eqref{eq09}, the true covariance exhibits the following decomposition:
\begin{align}\label{eq11}
	\Cv^{(i)}_y	=\mathcal{H}_i(\Lv^{(i)})\left( 	\mathcal{H}_i(\Lv^{(i)})\right) ^T=\Vv^{(i)}\Lambdav^{(i)} (\Vv^{(i)})^T,
\end{align}
where $\Lambdav^{(i)}=\diag([\lambda_1^{(i)},\cdots,\lambda_n^{(i)}])$ with $\{\lambda_{j}^{(i)}\}_{j=1}^n$ sorting the frequency responses $\{(\tilde h_{k}^{(i)})^2\}_{k=1}^n$ in descending order. Assumption \ref{ass2} ensures that both $\Lambdav^{(1)}$ and $\Lambdav^{(2)}$ have distinct diagonal entries, and they are aligned in the same order.  We represent the eigendecomposition of $\widehat \Cv_y^{(i)}$ as:
\begin{align}\label{eqeigen}
	\widehat \Cv_y^{(i)}=\Uv^{(i)}\widehat \Lambdav^{(i)}(\Uv^{(i)})^T,
\end{align}
where $\Uv^{(i)}$ is the sample eigenvector matrix, and $\widehat\Lambdav^{(i)}$ is the diagonal matrix with diagonal elements sorted in descending order as $\hat\lambda_1^{(i)}\geq \hat\lambda_2^{(i)}\geq\cdots\geq \hat\lambda_n^{(i)}\geq 0$.

As discussed in Section \ref{sec_2c}, the fact that the observations are generated by different graph filters makes it inappropriate to directly extend the conventional graph matching approach in \eqref{eq02} by replacing $\Lv^{(i)}$ with  $\widehat \Cv^{(i)}_y$. 
Note that  $\Uv^{(i)}$ and $\widehat \Lambdav^{(i)}$ in \eqref{eqeigen} provide approximations to  $\Vv^{(i)}$ and $ \Lambdav^{(i)}$ in \eqref{eq11}, respectively. Accordingly, we approximate $\overline\Vv^{(i)}$ in \eqref{eq05} by $\overline\Uv^{(i)}$, where $[\overline\Uv^{(i)}]_{k,l}=|u_{kl}^{(i)}|,\forall k,l$.
This motivates us to extend the conventional spectral method in \eqref{eq05} to the blind case by:
\begin{align}\label{eq12}
	\widehat \Pv=\argmax_{\Pv\in\mathcal{P}_n} \tr\left(\Pv^T\overline\Uv^{(1)}_K(\overline\Uv^{(2)}_K)^T\right),
\end{align}
where $K\leq n$ is a predefined hyper-parameter, and $\overline\Uv^{(i)}_K\in\Real_{+}^{n\times K}$ is the submatrix of $\overline\Uv^{(i)}$ containing  the left $K$ columns of  $\overline\Uv^{(i)}$. Similar to \eqref{eq05}, we use the absolute values of the sample eigenvectors to compute the node matching. This approach effectively resolves the unknown sign ambiguities inherent in the eigendecomposition; see \eqref{eq04} and \eqref{eq5a}.
Different from \eqref{eq05} that uses all the eigenvectors, we employ the reduced $K$-dimensional eigen-subspace in \eqref{eq12} to avoid large perturbations caused by sampling error and signal noise. The method for choosing $K$ can be found in Section \ref{sec_4d}.

\begin{algorithm}[!t]
	\caption{The blind graph matching algorithm.}
	\label{algall}
	%	\label{algsca}
	\begin{algorithmic}[1]
		\STATE\textbf{Input:}  $n$, $K$, $\sigma^2$, and $\{\yv^{(i)}_m\}_{m=1}^M,i=1,2$.
		%		\STATE\textbf{Initialization:}   $\widehat \Pv=\bf 0$. 
		\STATE Compute the sample covariance matrices $\widehat \Cv^{(i)}_y$ by \eqref{eq10};\\
		\STATE Compute the eigendecomposition of $\widehat \Cv^{(i)}_y$ by \eqref{eqeigen};\\
		%\STATE For any $\Pv\in\mathcal{S}_n$, check the identifiability by \eqref{eq14};\\ 
		%\STATE \textbf{if} the matching is identifiable \textbf{then}\\
		\STATE Compute $\widehat \Pv$ by solving \eqref{eq12};\\
		%		\STATE\textbf{end for} \\
		\STATE	\textbf{Output:} {$\widehat\Pv$.}
	\end{algorithmic}
\end{algorithm}
We summarize the proposed blind matching algorithm in Algorithm \ref{algall}. Step 4 of Algorithm \ref{algall} requires solving the linear assignment problem in \eqref{eq12}, whose solution is presented in what follows.
\subsection{Solution to \eqref{eq12}}
To solve the linear assignment problem in \eqref{eq12}, one can use the Hungarian method \cite{Hungarian}, as previously employed in \cite{GMP_eign1}. Alternatively, we adopt a faster greedy approach with comparable accuracy \cite{1544893,lu2016fast}. Specifically, we iteratively select the row and column of the largest uncovered entry in $\overline\Uv_K^{(1)}(\overline\Uv_K^{(2)})^T$ until all entries are covered. This leads to an $n$-iteration greedy method as shown in Algorithm \ref{alg1}.
	\begin{algorithm}[!t]
		\caption{The greedy method for solving \eqref{eq12}.}
		\label{alg1}
		%	\label{algsca}
		\begin{algorithmic}[1]
			\STATE\textbf{Input:}  
			$\Gv=\overline\Uv_K^{(1)}(\overline\Uv_K^{(2)})^T$.
			\STATE\textbf{Initialization:}   $\widehat \Pv=\bf 0$. 
			\STATE \textbf{for} $\mbox{iter}=1,2,\cdots,n$\\
			\STATE~~Find $(i,j)=\argmax_{(i,j)} [\Gv]_{ij}$;
			\STATE~~Set $[\widehat\Pv]_{ij}=1$;\\
			\STATE~~Delete the $i$-th row and the $j$-th column of $\Gv$.
			\STATE\textbf{end for} \\
			\STATE	\textbf{Output:} {$\widehat\Pv$.}
		\end{algorithmic}
	\end{algorithm}

It is worth noting that the Hungarian and greedy methods differ in terms of both accuracy and computational complexity. The Hungarian method provides an optimal solution to \eqref{eq12}, while the greedy method is generally sub-optimal.
On the other hand, the computational complexity of the Hungarian is $\mathcal O(n^3)$, while the greedy method runs faster with a complexity of $\mathcal{O}(n^2(\log n+K))$. Based on the insights from our practical implementation experience, the Hungarian method fares better for graphs with small-to-intermediate sizes. When the graph size is large, \eg $n\geq 100$, the efficient greedy method is preferable.

\remark[Impact of graph symmetry]{\label{remark1}
We emphasize that the solution to graph matching, even in the absence of signal noise, is not necessarily unique. This is the case when at least one of the graphs contains symmetric structures, such as cycles or symmetric trees. In such cases, node permutations are subject to non-trivial graph automorphisms, which complicates the matching accuracy evaluation for symmetric nodes.

To facilitate matching performance evaluation, it is useful to assess the \emph{identifiability} of the graph matching problem by identifying the symmetric nodes. Given the graph Laplacian matrices, one can verify if $\Lv^{(i)}\neq\Pv^T\Lv^{(i)}\Pv,i=1,2,$ for any $\Pv\neq \Iv_n,\Pv\in \mathcal{P}_n$. However, since $|\mathcal{P}_n|=n!$, this approach is computationally prohibitive for large $n$. A more feasible alternative is to only find the symmetric nodes subject to a single swap. Specifically, a node $k$ in graph $\mathcal{G}_i$ is asymmetric \emph{only if} $\Lv^{(i)}\neq \Pv_{k,j}^T\Lv^{(i)}\Pv_{k,j}$, or equivalently, $\Vv^{(i)}\neq \Pv_{k,j}^T\Vv^{(i)}$, for all $j\neq k$, where $\Pv_{k,j}$ is the swapping matrix that swaps the $k$-th and $j$-th columns of $\Iv_n$.
When the graph Laplacian matrices are unknown, the single-swap symmetric nodes can be approximately detected by substituting $\Vv^{(i)}$ with $\Uv^{(i)}$ from \eqref{eqeigen}.

We note that this method is a necessary, but not sufficient, condition to confirm asymmetry. Therefore, it only serves as a preliminary heuristic to filter out part of symmetric nodes when evaluating matching algorithms. For a more robust assessment, one can refer to the sufficient conditions for symmetric graphs in \cite{symmetry02,10.1093/imaiai/iav002} to identify a subset of asymmetric nodes.
}
\section{Performance Analysis and Eigenvector Selection}\label{sec4}
In this section, we analyze the performance of the blind graph matching approach in \eqref{eq12} by quantifying the impacts of the signal sampling size, observation noise,  and graph filters. Moreover, based on the analytical result, we propose a method to choose the system parameter $K$ in \eqref{eq12}. 

Throughout this section, we assume that the graphs $\mathcal{G}_1$ and $\mathcal{G}_2$ are asymmetric and exactly isomorphic. This implies that the optimal solution to the error-free spectral method \eqref{eq05} is unique and also optimal to \eqref{eq02}, \ie $\Pv^{\star}=\Pv^{\star\star}$. 
Note that the eigenbases of the sample covariances $\Uv^{(i)}$ and $\widehat \Lambdav^{(i)}$ are noisy estimates of those of  the true covariance matrices $\Vv^{(i)}$ and $\Lambdav^{(i)}$. Consequently, the permutation $\widehat\Pv$ obtained from the blind problem in \eqref{eq12} is generally sub-optimal compared with $\Pv^\star$. We bound the `sub-optimality' of $\widehat \Pv$ to $\Pv^\star$ by first analyzing the perturbations in $\Uv^{(i)}$ and $\widehat \Lambdav^{(i)}$. 

\subsection{Error in Sample Eigenvalues}
Recall from \eqref{eq11} and \eqref{eqeigen} that $\hat \lambda_k^{(i)}$ and $\lambda_k^{(i)}$ are the $k$-th largest eigenvalues of the sample covariance $\widehat \Cv_y^{(i)}$ and the true covariance $ \Cv_y^{(i)}$, respectively.
%From \eqref{eq09}, we have $\lambda_{o(k)}^{(i)}=(\tilde h_k^{(i)})^2$ with the frequency response ordered by $o(\cdot)$. 
Accordingly, $\{\hat \lambda_k^{(i)}\}$  can be regarded as shuffled sample estimates of the filter frequency response squares.

As shown in Section \ref{sec_2c}, in order to obtain an accurate permutation, it is necessary to align the eigenvectors w.r.t. the two graphs according to the same order of eigenvalues. This is guaranteed for the error-free setup in  \eqref{eq05} by Assumption \ref{ass2}. However, in the blind problem, $\hat \lambda_k^{(i)}$ is a perturbed estimate of $\lambda_k^{(i)}$ due to the finite number of signal samples and observation noise. As a result, the order of the frequency responses may not be preserved in the sample eigenvalues  $\{\hat \lambda_k^{(i)}\}$ if the perturbation is substantial.  

By utilizing the perturbation analysis on sample covariances, we analyze the influence of the sampling size $M$ and the observation noise variance $\sigma^2$ to the sample eigenvalues as follows. By noting that $\E[\widehat \Cv_y^{(i)}]=\Cv_y^{(i)}$, the result in \cite[Corollary 4.2]{Eigen_perturbation} provides the following concentration bound on the perturbation in the sample eigenvalue.
\lemma[cf. \cite{Eigen_perturbation} ]{\label{lemma2}Suppose $\{\yv_m^{(i)}\}_{m=1}^M$ are independent and identically distributed (i.i.d.) with a finite fourth-moment. For any $t>0$ and fixed $k\in[n]$, we have
	\begin{align}\label{eq19}
		\Pr\left(|	\hat \lambda_k^{(i)}-  \lambda_k^{(i)}|\geq t\right)\leq \frac{\kappa_k^{(i)}}{Mt^2},
	\end{align}
	where $\kappa_k^{(i)}=\E[\norm{\yv_m^{(i)}(\yv_m^{(i)})^T\vv_k^{(i)}}_2^2]-\lambda_k^{(i)}\leq \E[\norm{\yv_m^{(i)}}_2^4]$.
	
}
The conditions in Lemma \ref{lemma2} can be satisfied by sub-Gaussian signals and large $n$. 
Lemma \ref{lemma2} shows that the perturbation in  the sample eigenvalue is small with high probability when $M$ is large. Based on Lemma \ref{lemma2}, we show that $\{\hat \lambda_k^{(i)}\}$ have the same alignment order as $\{\lambda_k^{(i)}\}$ under such conditions. 
For $\lambda_1^{(i)}\geq \lambda_2^{(i)}\geq\cdots\geq \lambda_n^{(i)}\geq 0$, denote the spectral gap w.r.t. $\lambda_k^{(i)}$ by $\delta_k^{(i)}=\min\{\lambda_k^{(i)}-\lambda_{k+1}^{(i)},\lambda_{k-1}^{(i)}-\lambda_{k}^{(i)}\}$, where we define $\lambda_0^{(i)}=\infty$ and $\lambda_{n+1}^{(i)}=-\infty$. 
Specifically, to ensure $\{\hat \lambda_k^{(i)}\}$  aligned with $\{\lambda_k^{(i)}\}$, it is sufficient to have $|	\hat \lambda_k^{(i)}-  \lambda_k^{(i)}|<\delta_k^{(i)}/2$ for $\forall k\in[n]$. The following proposition characterizes the condition of aligned eigenvalues. 
\proposition{\label{pro1}Suppose the conditions in Lemma \ref{lemma2} hold. %Moreover, suppose the noise satisfies $\sigma^2\leq \frac{1}{2}\min_{k}\{\delta_k^{(1)},\delta_k^{(2)}\}$.
	For any fixed $k\in[n]$, with probability at least $1-\frac{4\kappa_k^{(i)}}{M({\delta_k^{(i)}})^2}$, one has that
	\begin{align}\label{eq18}
		|	\hat \lambda_k^{(i)}-  \lambda_k^{(i)}|<\frac{\delta_k^{(i)}}{2}.
	\end{align}
}
\begin{IEEEproof}
	% Applying the triangle inequality, we have $|	\hat \lambda_k^{(i)}-  \lambda_k^{(i)}-\sigma^2|\geq ||	\hat \lambda_k^{(i)}-  \lambda_k^{(i)}|-\sigma^2|$. Applying Lemma \ref{lemma2}, we have 
	% \begin{align}\label{eq19}
	% 	\Pr\left(|	\hat \lambda_k^{(i)}-  \lambda_k^{(i)}|< t+\sigma^2\right)\geq 1- \frac{\kappa_k^{(i)}}{Mt^2}.
	% \end{align}
	The result directly follows from Lemma \ref{lemma2} by setting $t=\delta_k^{(i)}/2$ in \eqref{eq19}.
 %leads to \eqref{eq18}.
\end{IEEEproof}

In summary, the accuracy of estimating the filter frequency responses from the sample covariance improves with a larger sampling size $M$ with probability increases with a larger spectral gap.
%. Moreover, when the noise variance $\sigma^2$ is small or the spectral gap is large, $\{\hat \lambda_k^{(i)}\}$  remains the same order as $\{\lambda_k^{(i)}\}$. 
The correct order of the sample eigenvalues is crucial to obtain precise graph matching, as detailed in the subsequent section.
\subsection{Analysis on Optimality Gap}
From \eqref{eq5b} and \eqref{eq05}, the optimal permutation $\Pv^\star$ maximizes the error-free matching objective in \eqref{eq05}. Denote $\overline \Vv_K^{(i)}$ as the submatrix of $\overline \Vv^{(i)}$ containing the left $K$ columns. In the noiseless setup with $\overline \Uv_K^{(i)}$ replaced by $\overline \Vv_K^{(i)}$ in \eqref{eq12}, we have
\begin{align}\label{eq_subspace}
	\tr(\Pv^T\overline\Vv_K^{(1)}(\overline\Vv_K^{(2)})^T)
	=&\sum_{j=1}^n\sum_{k=1}^K |v_{\pi(j)k}^{(1)}||v_{jk}^{(2)}|\nonumber\\
	=&\sum_{k=1}^K (\overline\vv_k^{(1)})^T(\Pv\overline\vv_k^{(2)})\nonumber\\
	{\leq}&\sum_{k=1}^K \norm{\overline\vv_k^{(1)}}_2\norm{\Pv\overline\vv_k^{(2)}}_2=K,
\end{align}
where the equality holds if the two graphs are isomorphic and $\Pv=\Pv^\star$. In other words, $\Pv^\star$ also maximizes \eqref{eq_subspace} for any $K\leq n$.
%with the maximum objective equal to $ \tr(	(\Pv^\star)^T\overline\Vv_K^{(1)}(\overline\Vv_K^{(2)})^T)=K$. 
In contrast, the solution from the blind graph matching $\widehat\Pv$ is sub-optimal to \eqref{eq_subspace}. To evaluate the difference between $\widehat\Pv$ and $\Pv^\star$,  we characterize the `optimality gap' of $\widehat \Pv$ to \eqref{eq_subspace} by bounding the objective difference $K-\tr(	\widehat\Pv^T\overline\Vv_K^{(1)}(\overline\Vv_K^{(2)})^T)$. To this end, we assume in this subsection that $|	\hat \lambda_k^{(i)}-  \lambda_k^{(i)}|<\delta_k^{(i)}/2$ holds for $\forall i,k$. This condition can be achieved with probability $1-\mathcal{O}(\frac 1 M)$ as shown in Proposition \ref{pro1}. The next result follows.
\proposition{\label{pro2}Suppose the following conditions hold:
	\begin{enumerate}[label=(\roman*)]
		\item $\mathcal{G}_1$ and $\mathcal{G}_2$ are isomorphic;
		\item Assumption \ref{ass2} holds;
		\item $|	\hat \lambda_k^{(i)}-  \lambda_k^{(i)}|<\delta_k^{(i)}/2$ holds for $i=1,2$ and $\forall k\in[K]$.
	\end{enumerate}
	Then, the inequality in \eqref{eq21} shown on top of next page holds,	where:
	\begin{align}
		\bm \Delta^{(i)}&\triangleq \widehat\Cv_y^{(i)}-\Cv_y^{(i)},
	\end{align}
	and the minimum spectral gap of the two graph filters with $\delta_k^{(i)}$ defined in Proposition \ref{pro1} is referred to as:
	\begin{align}
		\delta_{\min,K}&\triangleq\min_{1\leq k\leq K}\{\delta_k^{(1)},\delta_k^{(2)}\}.
	\end{align}
	\begin{figure*}
		\begin{align}\label{eq21}
			&~~~~K-\tr(	\widehat\Pv^T\overline\Vv_K^{(1)}(\overline\Vv_K^{(2)})^T)
			\leq	\frac{2n\sqrt{2K}}{\delta_{\min,K}}\left( \norm{\Deltav^{(1)}}_2+\norm{\Deltav^{(2)}}_2\right)+\frac{4K}{(\delta_{\min,K})^2}\left( \norm{\Deltav^{(1)}}_2^2+\norm{\Deltav^{(2)}}_2^2+2(n+1)\norm{\Deltav^{(1)}}_2\norm{\Deltav^{(2)}}_2\right),\\
   &~~~~\big\lVert{ \frac{1}{M}\sum_{m=1}^M\tilde\yv^{(i)}_m(\tilde\yv^{(i)}_m)^T-\Cv_y^{(i)}}\big\rVert_2\leq  C_i\left(\sqrt{\frac{Y^2n\ln(N/t)}{M}}+\frac{Y^2n\ln(N/t)}{M}\right),\label{lemma2_eq1}\\
 &  ~~~~ \big\lVert{ \frac{1}{M}\sum_{m=1}^M\wv^{(i)}_m(\wv^{(i)}_m)^T-\sigma^2\Iv_n}\big\rVert_2\leq \sigma^2C_i\left(\sqrt{\frac{n\ln(N/t)}{M}}+\frac{n\ln(N/t)}{M}\right).\label{lemma2_eq2}
		\end{align}
		\hrulefill
	\end{figure*}
	
}
\begin{IEEEproof}
	See Appendix \ref{appb}.
\end{IEEEproof}

According to Proposition \ref{pro2}, the optimality gap is small when 1) the minimum spectral gap (determined by the graph filter frequency responses) is large, and 2) the distance between $\widehat\Cv_y^{(i)}$ and $\Cv_y^{(i)}$ is small. As shown later in Section \ref{sec_4d}, we can select $K$ to ensure the ratio $\frac{\sqrt{K}}{\delta_{\min,K}}$ remains approximately constant. Combining Propositions \ref{pro1} and \ref{pro2}, the optimality gap grows at a rate of $\mathcal{O}(n(\norm{\Deltav^{(1)}}_2+\norm{\Deltav^{(2)}}_2))$ with probability $1-\mathcal{O}(\frac 1 M)$.
The covariance estimation error $\norm{\Deltav^{(i)}}_2$ critically affects the bound in \eqref{eq21}, which captures the combined impact of the finite number of samples $M$ and the noise in the observed signals. By following \cite[Exercise  5.6.4]{HDimPro} and \cite[Lemma 1]{GSP1}, we have the following bound on $\norm{\Deltav^{(i)}}_2$.

\lemma[cf. \cite{HDimPro,GSP1}]{\label{lemma5}Suppose $\{\yv_m^{(i)}\}_{m=1}^M$ are i.i.d. and uniformly bounded above almost surely with $\norm{\yv_m^{(i)}}_2\leq Y$.
%Let $r_i=\tr(\Cv_y^{(i)})/\norm{\Cv_y^{(i)}}_2\leq n$ be the effective rank of $\Cv_y^{(i)}$. 
For any $t>0$ and $i=1,2,$ there exists a constant $M_0\geq \max\{1,Y^2\}n\ln(n/t)$ and an absolute constant $C_i>0$ independent to $M,n,\sigma^2,$ and $t$ such that, for any $M\geq M_0$ and with probability at least $1-2t$,
	\begin{align}\label{eq22}
		\norm{\Deltav^{(i)}}_2\leq (\sigma^2+Y)C_i\sqrt{\frac{n\ln(n/t)}{M}}.
	\end{align}
}
\begin{IEEEproof}
    Define  $\tilde\yv^{(i)}_m=\yv^{(i)}_m-\wv^{(i)}_m$. For sufficiently large $M$, we have
    		\begin{align}
				\norm{\Deltav^{(i)}}_2=\norm{	\widehat \Cv^{(i)}_y-\Cv_y^{(i)}}_2&\leq \big
				\lVert{ \frac{1}{M}\sum_{m=1}^M\tilde\yv^{(i)}_m(\tilde\yv^{(i)}_m)^T-\Cv_y^{(i)}}\big\rVert_2\nonumber\\
    &+\big
				\lVert{ \frac{1}{M}\sum_{m=1}^M\wv^{(i)}_m(\wv^{(i)}_m)^T-\sigma^2\Iv_n}\big\rVert_2,
			\end{align}
   where the inequality follows from triangle inequality and $\sum_{m=1}^M\tilde\yv^{(i)}_m(\wv^{(i)}_m)^T\to 0$ as $M\to \infty$. Applying the results in \cite[Exercise  5.6.4]{HDimPro} and \cite[Lemma 1]{GSP1}, we establish bounds for the two terms on the right-hand side, each with a probability of at least $1-t$ in \eqref{lemma2_eq1} and \eqref{lemma2_eq2}, as shown on top of the next page.
   To finalize the proof, we set $M_0\geq \max\{1,Y^2\}n\ln(n/t)$. Applying the union probability bound and simplifying the resulting bound complete the proof.
\end{IEEEproof}

By applying Lemma \ref{lemma5} to \eqref{eq21}, it follows that the optimality gap in \eqref{eq21} increases at the rate of $\mathcal{O}\left(\sqrt{\frac{n^3\log n}{M}}\right)$ when $M\geq M_0\gtrsim n\log n$. Specifically, when $M$ scales proportionally to $n\log n$, the optimality gap exhibits a growth of $\mathcal{O}(n)$. In the subsequent section, we shall demonstrate that this choice of $M$, combined with establishing an upper bound on the noise variance $\sigma^2$, is adequate to ensure a diminishing matching error probability.

%This indicates that accurate blind graph matching can be achieved with $M\gg 1$ and $\sigma^2\ll 1$.
\subsection{Analysis on Matching Error Probability}\label{sec_4c}
Besides analyzing the optimality gap w.r.t. the matching objective, we further investigate in this section the probability of $\widehat \Pv$ making incorrect node matching compared with $\Pv^\star$. To this end, we derive an upper bound on the probability of $\widehat \Pv\neq \Pv^\star$.

To proceed, denote the node mapping functions w.r.t. $\widehat \Pv$ and $\Pv^\star$ by $\hat \pi(\cdot)$ and $\pi^\star(\cdot)$, respectively.
The optimal objective value of \eqref{eq_subspace} can be represented as
\begin{align}\label{eq23}
	&\tr\left((\Pv^\star)^T\overline\Vv_K^{(1)}(\overline\Vv_K^{(2)})^T\right)=\sum_{j=1}^n \left[\overline\Vv_K^{(1)}(\overline\Vv_K^{(2)})^T\right]_{\pi^\star(j),j}.
\end{align}

Motivated by \eqref{eq23}, we denote the $(\pi^\star(j),j)$-th entry of $\overline\Vv_K^{(1)}(\overline\Vv_K^{(2)})^T$ by $c_j$ and denote the maximum value in the $j$-th column excluding $c_j$ by 
\begin{align}\label{eq23b}
\ell_j\triangleq\max_{l\neq \pi^\star(j)}\left[\overline\Vv_K^{(1)}(\overline\Vv_K^{(2)})^T\right]_{l,j}.
\end{align}  
Furthermore, we define 
\begin{align}\label{eq24}
	\rho\triangleq\min_{j\in[n]}\left(c_j-\ell_j\right).
\end{align}
Intuitively, $\rho$ quantifies the maximum spectral leakage from each correctly matched entry $c_j$ (or equivalently, the inner product of the two correctly matched row eigenvectors) to the mismatched entries.
It follows from the Cauchy–Schwarz inequality that $c_j$ and $\ell_j$ lie in the range of $[0,1]$ for $\forall j$, implying that $\rho\in[-1,1]$. Note that the value of $\rho$ is an intrinsic characteristic of the graph, that can be computed numerically for a specific graph matching problem given the graph Laplacian. When the Laplacian is unknown, the expressions that depend on $\rho$ are useful to shed light on trends. 
The next result characterizes the error probability of blind graph matching.
\proposition{\label{pro3}For any specific graph matching problem with $\rho$ given, suppose the following conditions hold: 
	\begin{enumerate}[label=(\roman*)]
		\item The conditions in Proposition \ref{pro2} and Lemma \ref{lemma5} hold;
		%		\item The problem \eqref{eq13} has a unique solution $\widehat \Pv$;
		\item $\rho>0$;
		\item The signal noise is bounded by 
		\begin{align}\label{eq_29}
			\sigma^2<\overline\sigma^2\triangleq  \frac{\rho \delta^2_{\min}(K)}{16K+8\sqrt{2K}\delta_{\min,K}}.
		\end{align}
	\end{enumerate}
	Then, there exists some constant $M_0>0$ such that for any $M\geq M_0$, we have
	\begin{align}\label{eq25}
		\Pr(\widehat \Pv\neq \Pv^\star)\leq 4n e^{-\frac{M}{nC}\left( \overline\sigma^2-\sigma^2\right)^2},
	\end{align}
	where $C$ is a constant independent to $M,n,K,\sigma^2,$ and $\rho$. 
}
\begin{IEEEproof}
	See Appendix \ref{appc}.
\end{IEEEproof}
Condition (ii) requires a positive spectral leakage $\rho$, \ie each $(\pi^\star(j),j)$-th entry of $\overline\Vv_K^{(1)}(\overline\Vv_K^{(2)})^T$ must be the unique largest among the entries of the $j$-th column. This condition is likely to be satisfied when $K=n$ since $c_j=1$ and $\ell_j\leq 1$. However, it may be violated when $K$ is small. Condition (iii) holds with a small signal noise or a large spectral gap of the graph filters. 

For fixed $M$ and $n$,  a large $\rho$ leads to a smaller error probability bound. This is because a larger $\rho $ makes the matching problem \eqref{eq05}  more robust against perturbations, resulting in greater tolerance on the signal noise and finite sampling size; see Condition (iii) and \eqref{eq25}. We note that the analytical result in \eqref{eq25} requires the knowledge of $\rho$. In the case of blind matching with unknown graph topologies, we can approximate $\rho$ by estimating its statistics using random graph models or by approximating $\overline \Vv_K$ with the sample eigenvectors $\overline\Uv_K$ in \eqref{eq23}.

Proposition \ref{pro3} suggests an exponential decay rate of the error probability w.r.t. $M$. Moreover, the bound supports the intuition that the blind matching error increases with $n$ as matching larger graphs is more susceptible to error. 	
In summary, Proposition 3 suggests that by selecting the sample size \(M\) to be proportional to \(n\log n\) and maintaining the noise variance bounded by \eqref{eq_29}, we can ensure \(\Pr(\widehat \Pv \neq \Pv^\star) \leq \epsilon\) for any constant \(\epsilon\in (0,1)\). These conditions, combined with the condition $\rho>0$, lead to a diminishing error probability.
%To summarize, Proposition \ref{pro3} implies that $\widehat\Pv$ is close to $\Pv^\star$ as  $M$ grows sufficiently large and $\sigma^2$ is sufficiently small.
\remark[Requirements on graph signals and graph filters]In the above analysis, we have imposed specific conditions on graph signals and filters. Proposition \ref{pro1} and Lemma \ref{lemma5} assume that graph signals are i.i.d. and bounded. Proposition \ref{pro3} stipulates that the signal noise has a bounded variance. Moreover, Propositions \ref{pro1}--\ref{pro3} necessitate sufficiently large spectral gaps in the covariance matrices, which consequently extends to a constraint on the spectral gaps of the graph filters.
Note that the graph filter, as defined in \eqref{eq06}, is influenced by the filter coefficients and the graph Laplacian. When the filter coefficients are fixed, achieving a substantial spectral gap typically requires strong edge connectivity within the graph \cite{gap1,gap2}. On the other hand, in cases where the graph exhibits strong community structures, as often observed in social networks, its Laplacian matrix typically has many small eigenvalues \cite{gap3}. In this case, certain low-pass filters can result in large spectral gaps, as they tend to amplify low frequencies and attenuate high frequencies. Here, we provide two practical applications to illustrate this.

\example[Diffusion Dynamics]{The diffusion model is widely adopted to represent temperatures within a geographical area and opinion dynamics in social networks; see \cite{LPGP_mag,GSP1,9757839} and the references therein. The associated graph filter is given by 
					$\mathcal{H}(\Lv)=(\Iv_n-\alpha\Lv)^{T_d}$ with $\alpha>0$ and $T_d>0$.
As shown in Example \ref{ex1}, this is a typical low-pass filter. Using the definition in \eqref{eq09}, the graph frequency response $\tilde h_k$ corresponding to the $k$-th eigenvalue of $\Lv$, denoted by $\gamma_k$, is given by $\tilde h_k=(1-\alpha\gamma_k)^{T_d}$. For any adjacent pair of eigenvalues of $\Lv$, denoted by $\gamma_{k+1}$ and $\gamma_k$, the spectral gap of the above filter can be approximated by using the first-order Taylor expansion, as $|\tilde h_{k+1}-\tilde h_k| \approx  \alpha T_d|1-\alpha \gamma_k|^{T_d-1} |\gamma_{k+1}-\gamma_{k}|$. When the eigenvalue $\gamma_k$ is small, we have $|1-\alpha \gamma_k|^{T_d-1} \approx 1$. The graph filter amplifies the spectral gap by a factor of approximately $\alpha T_d$. Consequently, we expect a large spectral gap in the graph filter when $\alpha$ and $T_d$ are large.
}
\example[Image Processing]{Low-pass smoothing filters are widely adopted in image processing applications. For example, the negative exponential filter $\mathcal{H}({\bf L}) =\beta e^{-\alpha {\bf L}}$, with $\beta,\alpha > 0$, has been applied for image smoothing; see, e.g., \cite{6122336}.
			Consider any pair of the adjacent eigenvalues of $\Lv$ denoted by $\gamma_{k+1}$ and $\gamma_k$. Using the first-order Taylor expansion, the frequency responses of the above filter satisfy that $|\tilde h_{k+1}-\tilde h_k| \approx \alpha  \beta e^{-\alpha \gamma_k} |\gamma_{k+1}-\gamma_{k}|$. When $\gamma_{k}$ is small, $e^{-\alpha \gamma_k}\approx 1$. We conclude that the graph filter scales the spectral gap by a factor of approximately $\alpha \beta $.
}

\remark[Requirements on the underlying graphs]{The matching algorithm requires the eigenvalues of the Laplacian matrix to be distinct. However, this condition may not hold for highly symmetric graphs. In addition, Condition (ii) in Proposition \ref{pro3} is intrinsically linked to the spectral characteristics of the underlying graphs, particularly imposing requirements on the edge connectivity and the correlation between them. For example, \cite{GMP_eigen7} investigated a metric similar to $\rho$ in \eqref{eq24}, albeit employing a variant of the eigenvector similarity measurement. It shows that for large ER graphs and large Gaussian models, this metric is likely to be positive with a high probability, given that both the edge connectivity and the correlation between the graphs are sufficiently strong. Inspired by this, we hypothesize that the condition $\rho > 0$ is also probable for large ER graphs and large Gaussian models. Consequently, this would lead to a diminishing error probability in Proposition \ref{pro3}, particularly under conditions of sufficiently small signal noise in \eqref{eq_29}.
%While a thorough analysis is a subject for future research, we will study this subject through numerical experiments, as detailed in Section \ref{sec5}.
}

\subsection{Method for Eigenvector Selection}\label{sec_4d}
The above analysis has shown the non-monotonic effect of $K$ to blind graph matching: On the one hand, for fixed $n$ and $M$, Propositions \ref{pro1}--\ref{pro3} show that the error in blind matching increases with the minimum spectral gap normalized by $\sqrt{K}$, \ie $\delta_{\min,K}/\sqrt{K}$, which is non-increasing with $K$. On the other hand, for fixed $M$ and $\sigma^2$, we wish for a larger $\rho$ to guarantee Conditions (ii) and (iii) of Proposition \ref{pro3}, implying that a large $K$ is better.
To balance these opposing effects, we propose a heuristic line-search method for determining $K$, as shown in Algorithm \ref{alg3}. In Step 3 of Algorithm \ref{alg3}, we ensure the eigenvectors in $\Uv_K^{(i)}$ correspond to non-zero eigenvalues. In Step 4, we stop including more eigenvectors when the normalized empirical spectral gap dramatically drops.

Step 4 of Algorithm \ref{alg3} ensures that the minimum empirical spectral gap is proportional to $\sqrt{K}$. Consequently, the ratio $\frac{\sqrt{K}}{\delta_{\min,K}}$ remains approximately constant, leading to a constant value for $\bar \sigma^2$ in \eqref{eq_29} that is independent to $K$ and $n$. This result, combined with \eqref{eq25}, implies that setting \( M \) proportional to \( n\log n \) is sufficient for achieving a constant error probability.
\begin{algorithm}[!t]
	\caption{The line search method for selecting $K$.}
	\label{alg3}
	%	\label{algsca}
	\begin{algorithmic}[1]
		\STATE\textbf{Input:} The sample eigenvalues $\widehat \Lambdav^{(i)}$ in \eqref{eqeigen}, and the threshold value $\varsigma$.
		\STATE\textbf{Initialization:}   $K=1$. 
		\STATE \textbf{for} $K\leq \min_{i=1,2}\{\rank(\widehat \Cv_y^{(i)})\}$\\
		\STATE~~\textbf{if} $\min_{i=1,2}\frac{\hat\lambda_{K}^{(i)}-\hat\lambda_{K+1}^{(i)}}{\sqrt{K}}\leq \varsigma$ \textbf{then}
		\STATE~~~~Stop;
		\STATE~~\textbf{else} 
		\STATE~~~~$K\leftarrow K+1$;\\
		\STATE\textbf{end for} \\
		\STATE	\textbf{Output:} {$K$.}
	\end{algorithmic}
\end{algorithm}

\section{Numerical Results}\label{sec5}
In this section, we evaluate the performance of the proposed graph matching algorithms by simulations.
\subsection{Experiment Setup}\label{sec5_a}
We carry out experiments on the following graphs:
\begin{itemize}
	\item \textbf{ER random graphs \cite{ERGraph}}: We generate the first unweighted graph $\mathcal{G}_1$ by the ER model with $n=50$ nodes and an edge probability of $0.4$. We study an exact graph matching task, where the second graph $\mathcal{G}_2$ is obtained by randomly shuffling the node labels of $\mathcal{G}_1$.
	\item \textbf{Barab\'{a}si-Albert (BA)  preferential attachment graphs \cite{BAGraphs}}: We use the BA model to generate an unweighted scale-free graph $\mathcal{G}_1$. Specifically, we start with $4$ initially placed nodes and generated a total of $50$ nodes, where each new node is attached to $4$ existing nodes selected randomly proportional to their degrees. The second graph $\mathcal{G}_2$ is obtained by random node shuffling. 
	\item \textbf{Gaussian model \cite{GMP_theory1}}: We study inexact graph matching over two weighted graphs by following \cite{GMP_theory1}. Specifically, we generate the adjacency matrix $\Av^{(1)}$ as a standard Gaussian matrix with $n=50$. The second adjacency matrix is computed by
	\begin{align}\label{eq28}
		\Av^{(2)}=(\Pv^\star)^T(\sqrt{1-\b^2}\Av^{(1)}+\b \Zv^\prime)\Pv^\star,
	\end{align}
	where $\Pv^\star$ is the true permutation matrix randomly drawn from $\mathcal{P}_n$, $\Zv^\prime$ is a standard Gaussian matrix independent to $\Av^{(1)}$, and $\b\in(0,1)$ controls the  correlation between $\Av^{(1)}$ and $\Av^{(2)}$. A smaller $\b$  indicates a larger correlation and more similar underlying graphs. 
	\item \textbf{Real social networks}: We consider two real-world social networks: 
%	1) the \emph{Zachary's karate club} network \cite{ZacharyClub}, modeling interactions between club members with $34$ nodes and $78$ edges; 
	1) the \emph{Highschool} network \cite{HighSchool}, modeling friendships between individuals with $70$ nodes and $366$ edges; 
%	and 2) the \emph{Facebook Page} network from \cite{Facepage} with nodes representing the TV show pages and edges representing mutual likes. We process the data by removing self-loops, symmetric subgraphs, and nodes with degree less than $10$, leading to a graph with $936$ nodes and $9,141$ edges. 
	and 2) one \emph{Facebook} ego network from \cite{Facebook}, capturing friendships between anonymous users with $348$ nodes and $2,866$ edges.
%	We preprocess the data by eliminating all symmetric nodes except for one, resulting in a graph with $307$ nodes and $2,411$ 	edges.
%	$4,039$ nodes and $88,234$ 
 For each network, 
% we construct an undirected and unweighted graph from the data. By following \cite{pedarsani2013bayesian}, we then 
we apply independent edge sampling to obtain two similar subgraphs $\mathcal{G}_1$ and $\mathcal{G}_2$ with a sampling probability of $0.98$. 
%	Finally, we delete the unconnected nodes in $\mathcal{G}_1$ and $\mathcal{G}_2$. 
\end{itemize}

Unless otherwise specified, we employ the opinion-dynamic model  \cite{LPGP_mag} for the two non-identical graph filters as $\mathcal{H}_1=(\Iv_n+0.1\Lv^{(1)})^{-1}$ and $\mathcal{H}_2=(\Iv_n+0.3\Lv^{(2)})^{-1}$. The filtered graph signals $\{\yv^{(i)}_m\}$ are computed by \eqref{eq_signal} with $\xv_m$ drawn from $\Norm({\bf 0},\Iv_n)$. We set the noise variance $\sigma^2$ to $0.01$ in  \eqref{eq_signal}.\footnote{We note that the bias correction in \eqref{eq10} is merely for facilitating the derivation of the error bound in Lemma \ref{lemma5}. For the practical implementation, subtracting a matrix proportional to $\Iv_n$ does not affect either the sample eigenvector computation in \eqref{eqeigen} or the sample spectral gaps in Algorithm \ref{alg3}. For consistency with the model in \eqref{eq10}, we assume the knowledge of $\sigma^2$ and use it in computing \eqref{eq10} for our algorithm. However, employing an estimate of $\sigma^2$ in \eqref{eq10}, or even omitting the term $-\sigma^2\Iv_n$, does not alter the numerical results presented in this section.} For the proposed method, we set %$\e=n/20$ in \eqref{eq14} and 
$\varsigma=(10n)^{-2}$ in Algorithm \ref{alg3}. The problem in \eqref{eq12} is solved by either the Hungarian method or the greedy method in Algorithm \ref{alg1}.

We compare the proposed blind matching method with the following two baselines:
\begin{itemize}
	\item \textbf{Error-free graph matching}: This method assumes that the graph Laplacian matrices $\Lv^{(1)}$ and $\Lv^{(2)}$ are perfectly known. We solve \eqref{eq05} by the Hungarian method to obtain the error-free matching when the graphs are exactly the same. {This baseline represents the best possible matching result achievable by our method within the spectral matching framework of \eqref{eq05}. However, it may be sub-optimal in cases where the two graphs are not precisely isomorphic.}
	\item \textbf{Two-step blind graph matching}: For the blind graph matching scenario, we compute the sample covariance of each graph and its eigendecomposition by \eqref{eq10} and \eqref{eqeigen}. Then, we estimate each graph Laplacian matrix using the topology inference approach in \cite[Eqs. (17) and (25)]{Top_inference}.
%	 with the CVXPY solver \cite{CVXPY}.
%	\footnote{We tune the hyper-parameter "$\e$" \cite[Eq. (17)]{Top_inference} according to the parameter tuning method in \cite[Section V-C]{Top_inference}.} 
	Denote the estimated Laplacian by $\widehat \Lv^{(i)},i=1,2.$  We compute the estimated adjacency matrix   $\widehat \Av^{(i)},i=1,2,$ as $[\widehat \Av^{(i)}]_{kk}=0$ for $k\in [n]$ and $[\widehat \Av^{(i)}]_{kl}=-[\widehat \Lv^{(i)}]_{kl}$ for $\forall k\neq l$. Finally, we employ the state-of-the-art spectral graph matching algorithm in \cite{GMP_eigen7} with the estimated adjacency matrices $\widehat \Av^{(i)}$.
\end{itemize}
We evaluate the performance of blind graph matching using two metrics: 1) the matching disagreement function in \eqref{eq1}, and 2) the average fraction of correctly matched node pairs, i.e.:
\begin{align}
	\frac{1}{n}	\E\left[\sum_{j=1}^n\mathbbm{1}_{\{\pi(j)=\pi^\star(j)\}}\right],
\end{align} 
where $\pi^\star(\cdot)$ is the true node matching function and $\mathbbm{1}$ is the indicator function, with $\mathbbm{1}_A=1$ if event $A$ is true and $\mathbbm{1}_A=0$ otherwise. 
We perform 50 Monte Carlo trials and report the average over all the trials unless otherwise specified.
\subsection{Results on Random Graph Models}
\begin{figure}[!t]
	\centering
	\includegraphics[width=2.8 in]{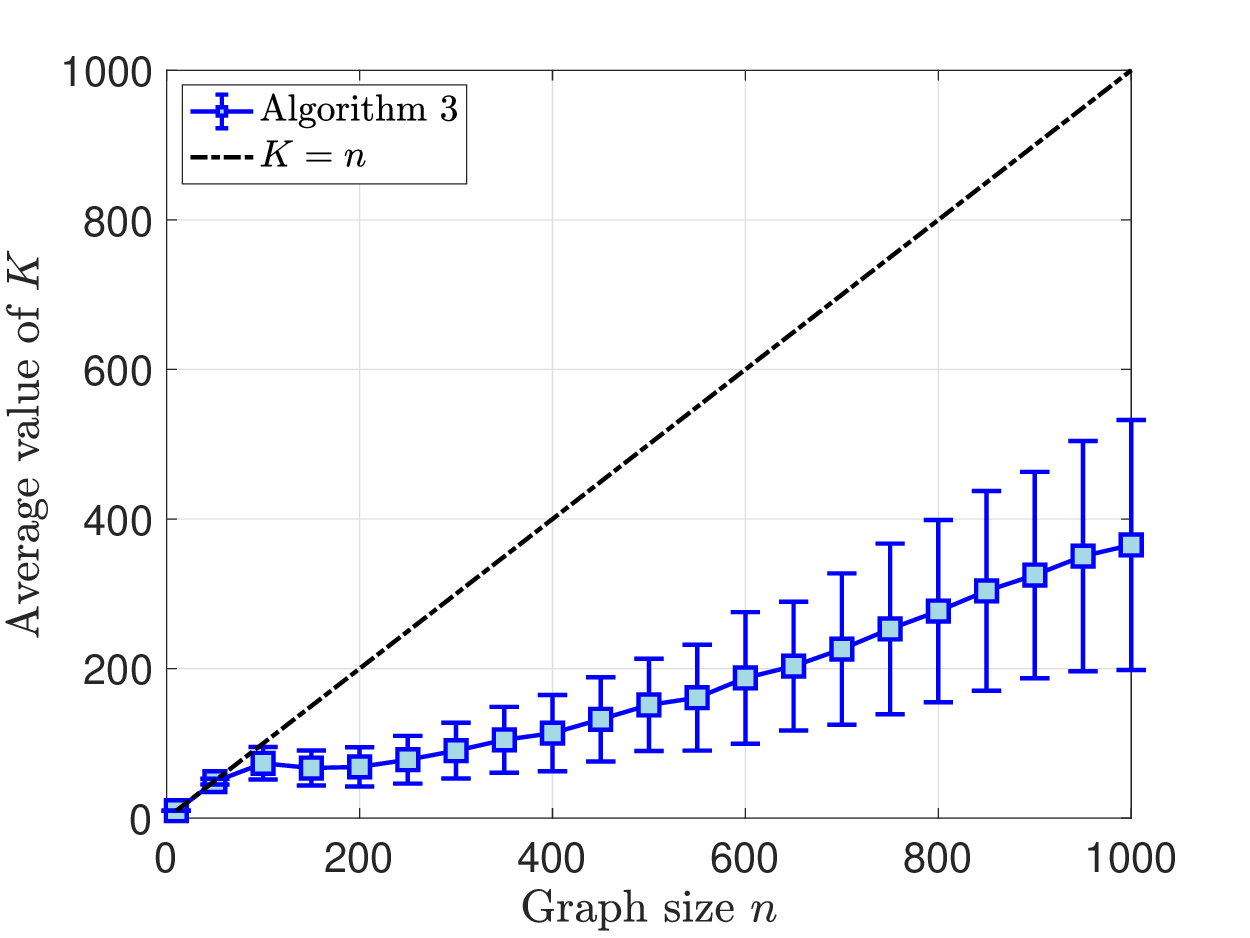}
	\caption{The average number of selected eigenvectors $K$ versus the graph size $n$. The vertical error bar at each point represents the empirical standard deviation in $500$ Monte Carlo trials.}
	\label{K}
\end{figure}

 First, we evaluate the effectiveness of the proposed eigenvector selection algorithm, \ie Algorithm \ref{alg3}. In Fig. \ref{K}, we simulate the ER graphs and analyze the average number of selected eigenvectors $K$ with a varying $n$. As Algorithm \ref{alg3} discards the eigenvectors associated with small spectral gaps, the number of required eigenvectors $K$ is much smaller than $n$. The result demonstrates that the proposed eigenvector selection scheme improves the computational efficiency of blind matching  by limiting a relatively small $K$ for large graphs.
\begin{figure}[!t]
	\centering
	\includegraphics[width=3.4 in]{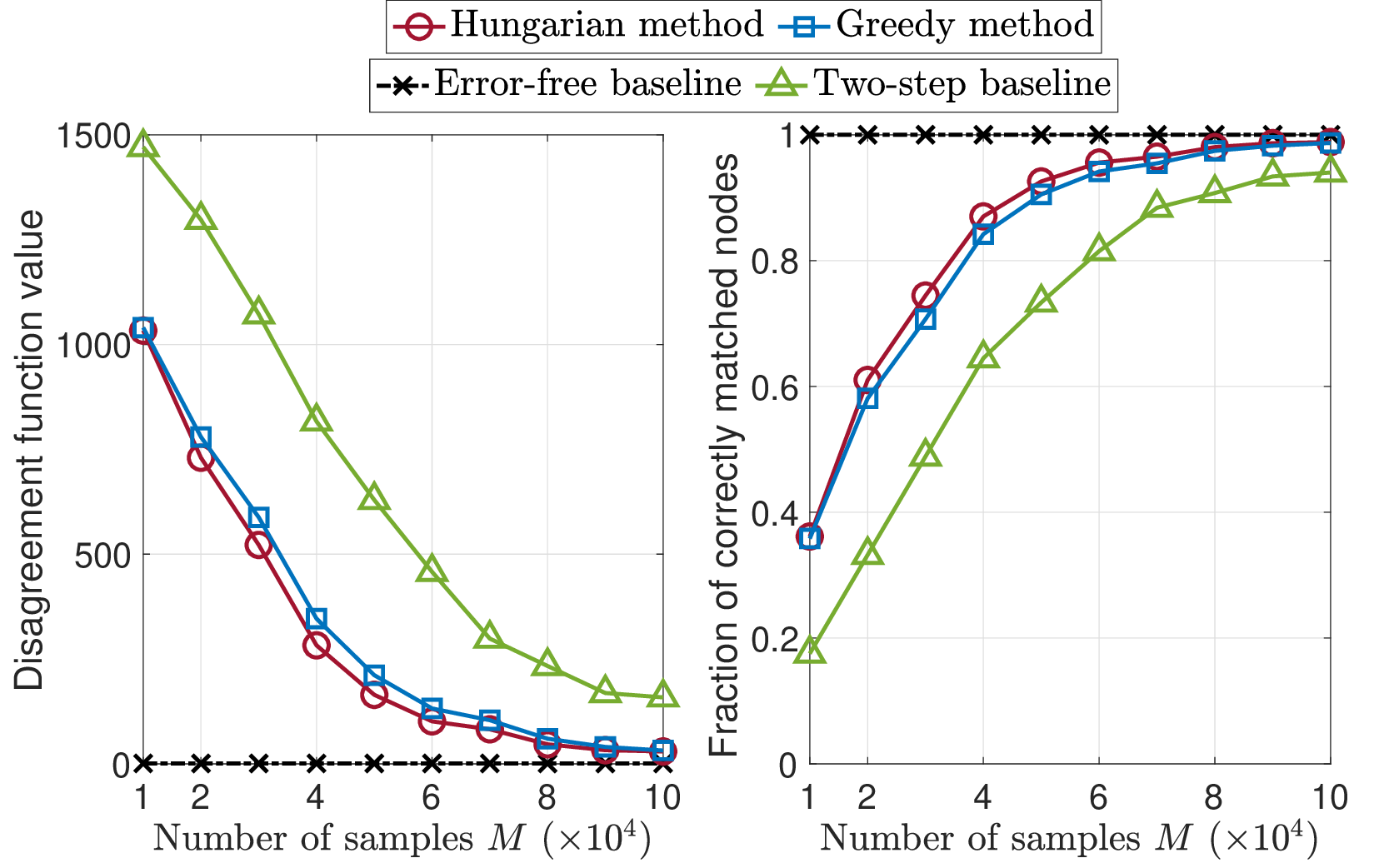}
	\caption{Performance of blind graph matching  versus the number of signal samples $M$ for the ER graphs.}
	\label{fig_M_ER}
\end{figure}
\begin{figure}[!t]
	\centering
	\includegraphics[width=3.4 in]{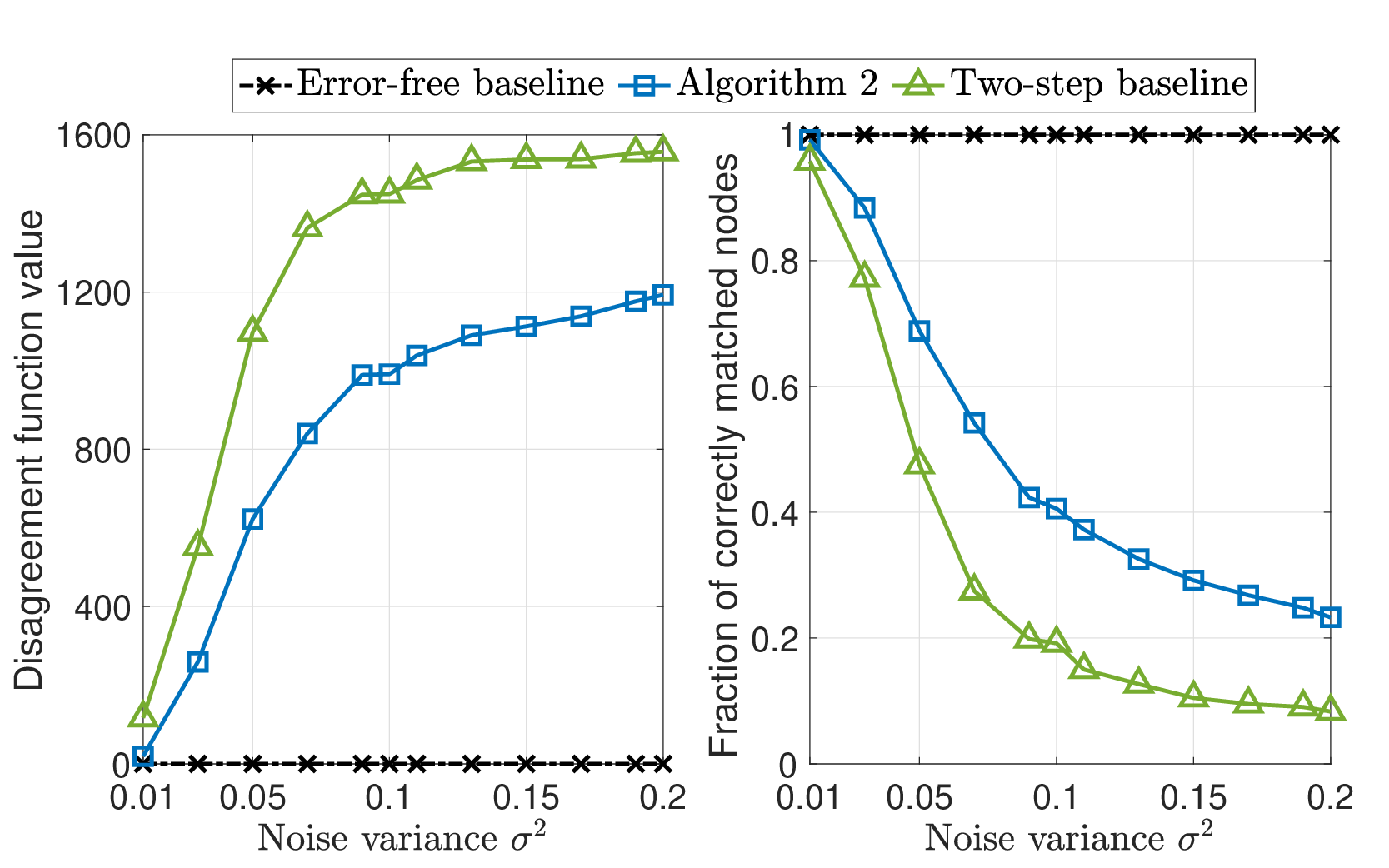}
	\caption{Matching performance  versus the signal noise variance $\sigma^2$ with $M=10^5$.}
	\label{fig_s_ER}
\end{figure}

We investigate the impact of the signal sampling size $M$ on the matching performance of the ER graphs. Fig. \ref{fig_M_ER} shows the disagreement function and the faction of correctly matched nodes for a varying $M$. The proposed methods and the two-step baseline use sample covariance matrices for matching; hence their accuracy increases with $M$. As expected, both the disagreement objective value and  the error rate decrease as $M$ increases, which aligns with the analysis in Section \ref{sec4}. In particular, the proposed method attains almost perfect graph matching with $M\geq 10^5$. On the other hand, the error-free baseline achieves perfect graph matching for this  exact matching experiment. We conclude from Fig. \ref{fig_M_ER} that the proposed approach outperforms the existing two-step baseline and achieves nearly perfect matching with a large $M$.

\begin{figure}[!t]
	\centering
	\includegraphics[width=3.4 in]{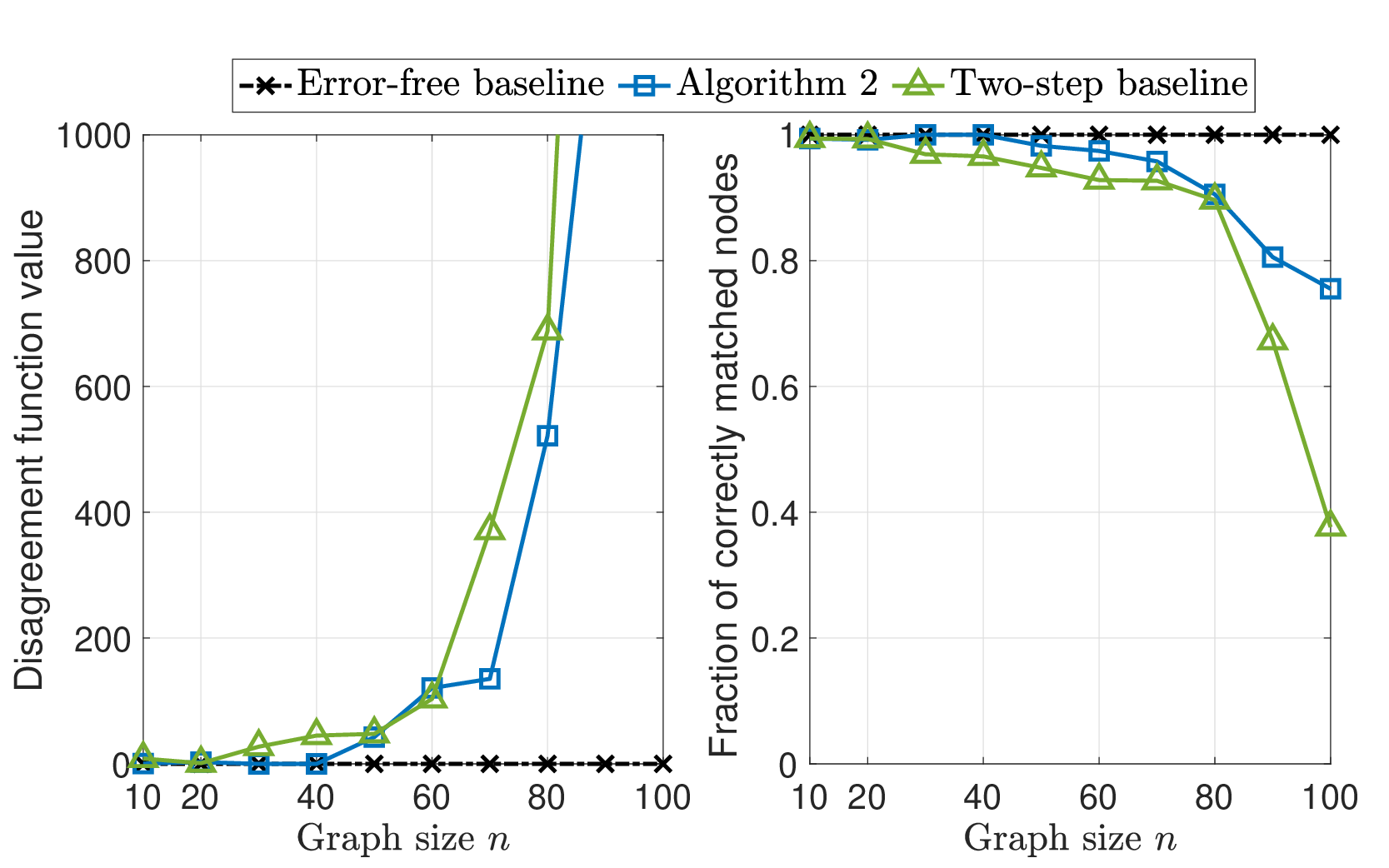}
	\caption{Matching performance under different graph sizes $n$, where we set $M=750\cdot n\ln n$ and $\sigma^2=0.01$.}
	\label{fig_n_ER}
\end{figure}

Next, we study the effect of signal noise in \eqref{eq_signal} on the matching performance of the ER graphs in Fig. \ref{fig_s_ER}. Since the Hungarian and greedy methods achieve similar performance in our approach, we choose to present only the results obtained using the greedy method, i.e., Algorithm \ref{alg1}, in the remaining simulations.
We adjust the noise variance $\sigma^2$ while fixing the signal sampling size at $M=10^5$. 
By way of comparison, the second largest and the smallest eigenvalues of the covariance matrix $\Cv_y^{(2)}$ are $\lambda^{(2)}_{2}\approx 0.075$ and $\lambda^{(2)}_{50}\approx 0.009$, respectively. It shows that a larger $\sigma^2$ leads to less accurate sample covariance matrices and greater perturbations in the eigendecomposition. Consequently,  the performance of blind matching deteriorates as $\sigma^2$ increases.
When the signal noise overwhelms the eigenvalues of the sample covariance, accurate graph matching becomes impossible even with a large number of signal samples. We see from Fig. \ref{fig_s_ER} that the proposed method outperforms the baseline in \cite{GMP_eigen7} at all levels of noise as it is more robust against signal noise.

Fig. \ref{fig_n_ER} illustrates the performance of our graph matching method for varying graph size $n$. According to Proposition \ref{pro3},  the error probability in blind graph matching grows at a rate of $\mathcal{O}(ne^{-M/n})$, suggesting that the signal sampling size $M$ should grow proportionally to $n\log n$. Motivated by this, we set $M=750n\ln n$ in Fig. \ref{fig_n_ER}. The result illustrates the robustness of our proposed method even for large $n$. In contrast, the two-step baseline is more prone to errors with large graphs, despite the increase in sample size.
\begin{figure}[!t]
	\centering
	\includegraphics[width=3.4 in]{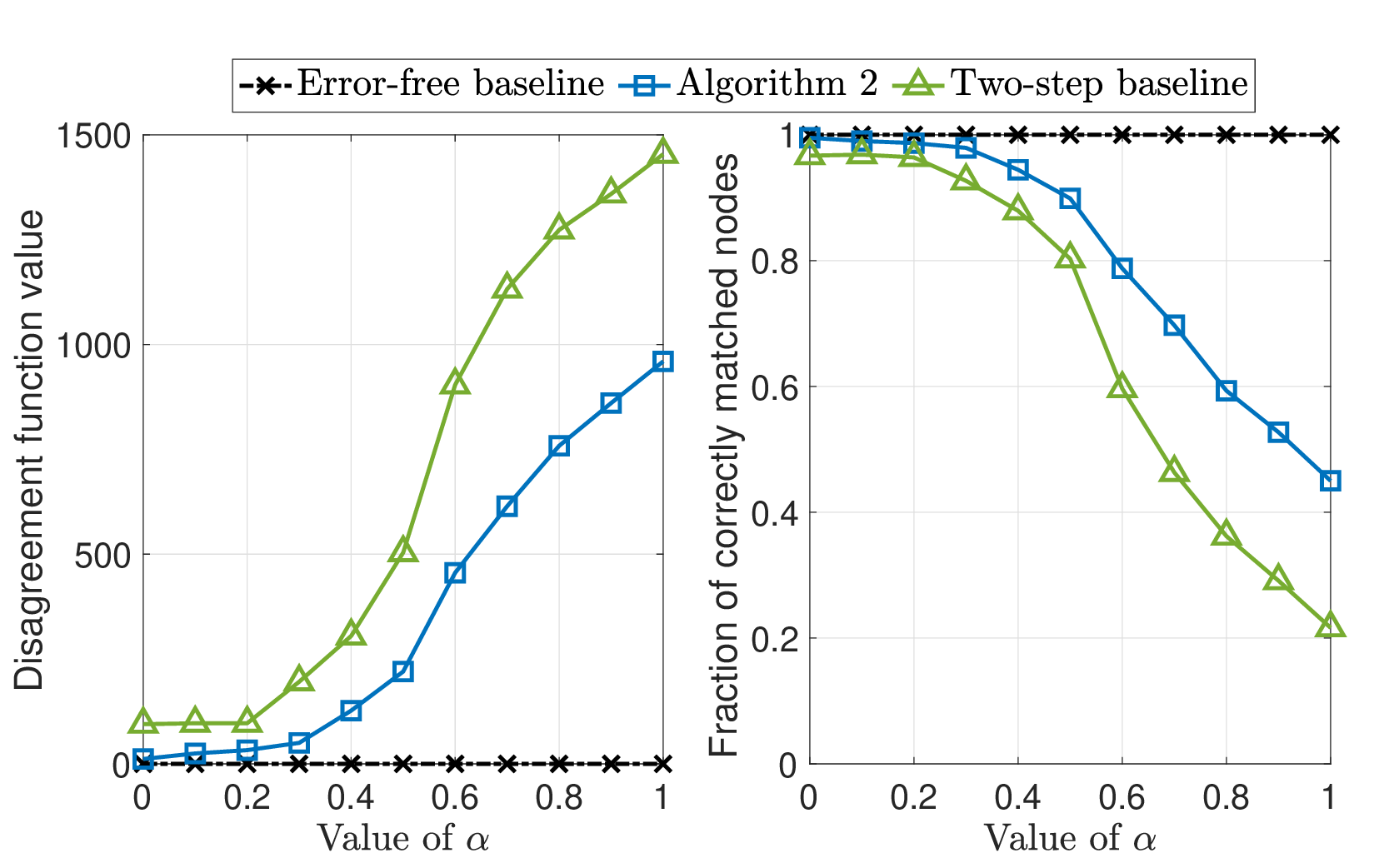}
	\caption{Graph matching with non-identical graph filters, where a larger $\a$ means more heterogeneous filters.}
	\label{fig_a_ER}
\end{figure}
Moreover, we set $\mathcal{H}_1=(\Iv_n+0.1\Lv^{(1)})^{-1}$ and $\mathcal{H}_2=(\Iv_n+(0.1+\a)\Lv^{(2)})^{-1}$ with $\a$ controlling the heterogeneity of the two graph filters. Fig. \ref{fig_a_ER} plots the performance of graph matching versus the value of $\a$. A larger $\a$ leads to a smaller spectral gap in the covariance matrix $\Cv_y^{(2)}$ and thus a larger error in blind graph matching. 
\begin{figure}[!t]
	\centering
	\includegraphics[width=3.4 in]{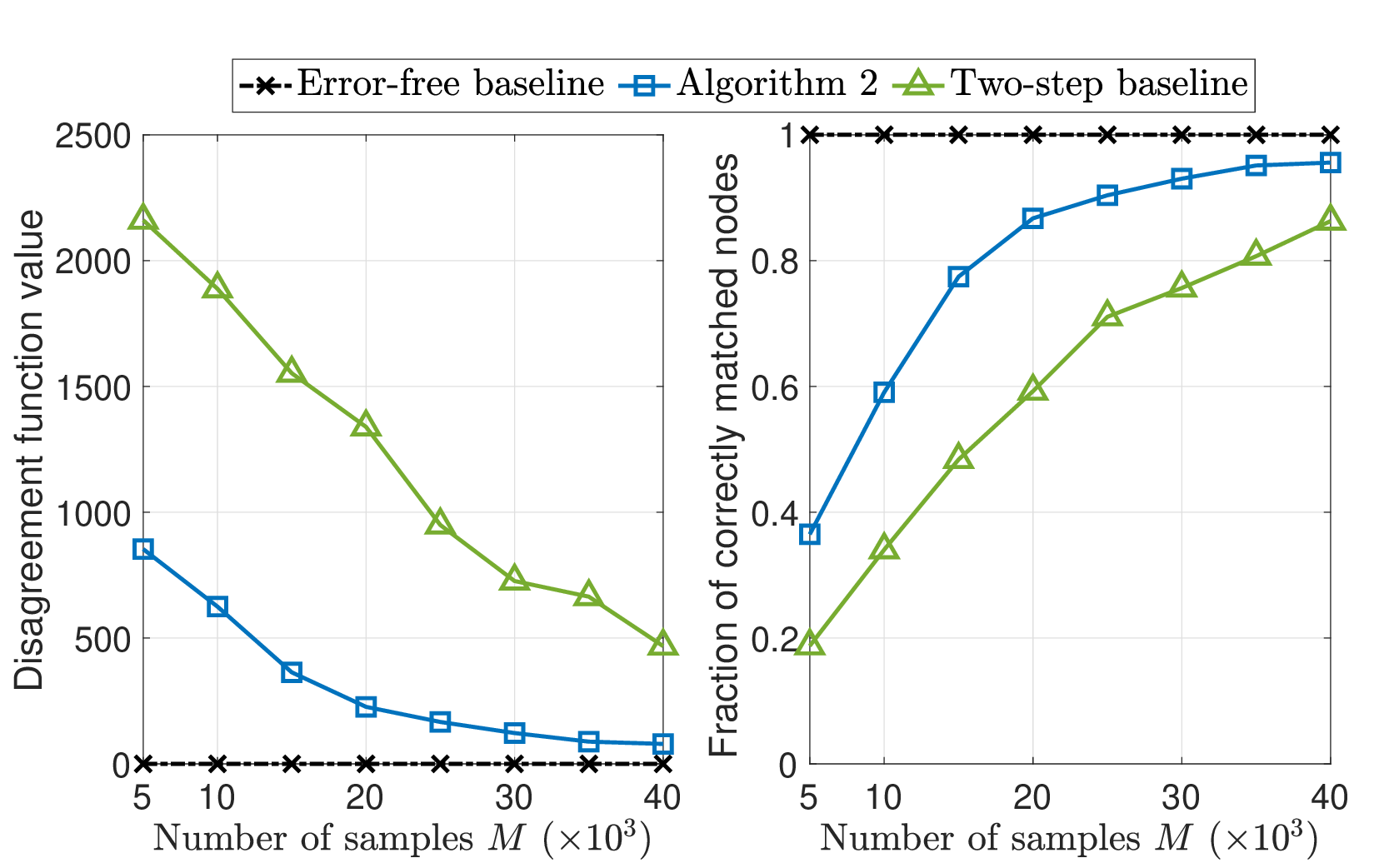}
	\caption{Graph matching performance for the BA graph model.}
	\label{fig_M_BA}
\end{figure}
\begin{figure}[!t]
	\centering
	\includegraphics[width=3.4 in]{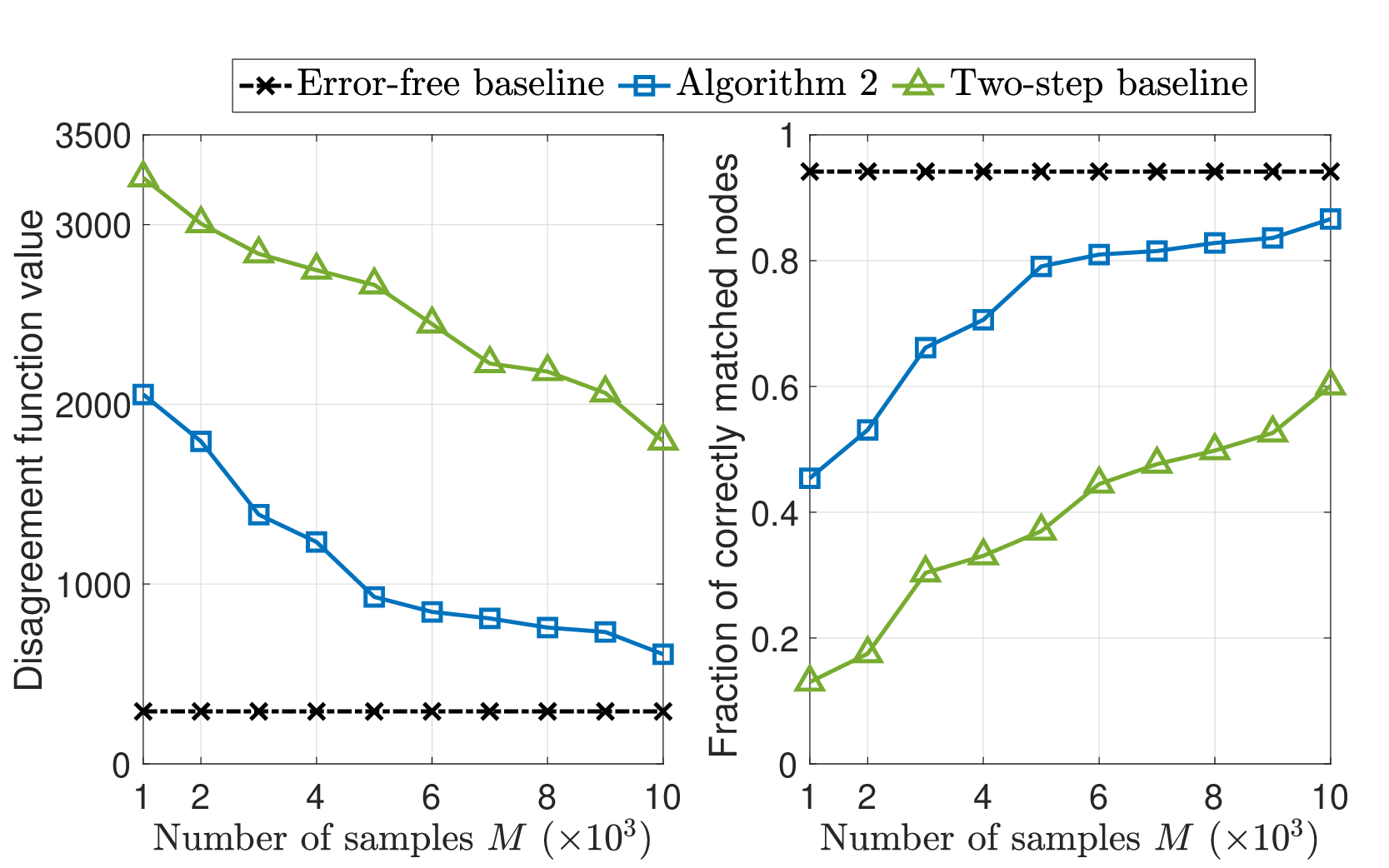}
	\caption{Inexact graph matching for the Gaussian model with $\b=0.1$ in \eqref{eq28}.}
	\label{fig_M_GW}
\end{figure}

In Figs. \ref{fig_M_BA} and \ref{fig_M_GW}, we investigate the performance of graph matching over the BA graph model and the Gaussian model, respectively. Here, we set $\sigma^2=0.01$, $n=50$, and $\a=0.2$ and vary the sample size $M$. Similar to Fig. \ref{fig_M_ER}, the proposed method achieves more accurate matching as $M$ increases. For the inexact matching on the Gaussian model, the error-free baseline in \eqref{eq05} is sub-optimal to \eqref{eq02}, leading to an imperfect matching in Fig. \ref{fig_M_GW}. 

\begin{figure}[!t]
	\centering
	\includegraphics[width=3.4 in]{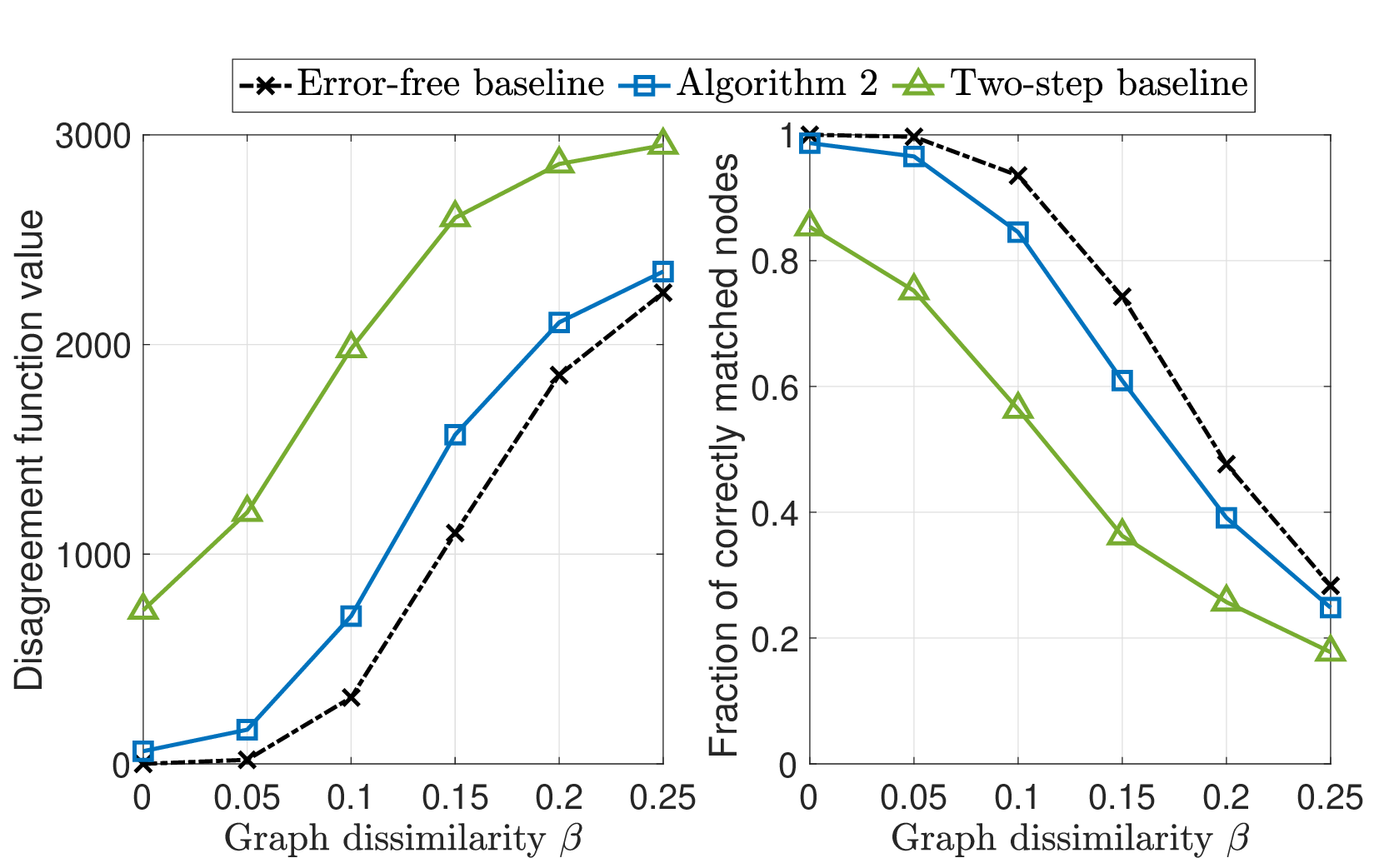}
	\caption{Performance of inexact graph matching on the Gaussian model versus the value of $\b$ with $M=10^4$.}
	\label{fig_b_GW}
\end{figure}

In Fig. \ref{fig_b_GW}, we vary the correlation parameter $\b$ in \eqref{eq28} to study inexact matching. A larger $\b$ means less correlation between their adjacency matrices and less similar underlying graphs $\mathcal{G}_1$ and $\mathcal{G}_2$. We see that all the algorithms exhibit larger errors as $\b$ increases. Our method achieves an accuracy close to the error-free baseline. In contrast, the two-step baseline is more prone to topology inference errors, resulting in inaccurate matching results.
\begin{figure}
								\centering
				\includegraphics[width=3.4 in]{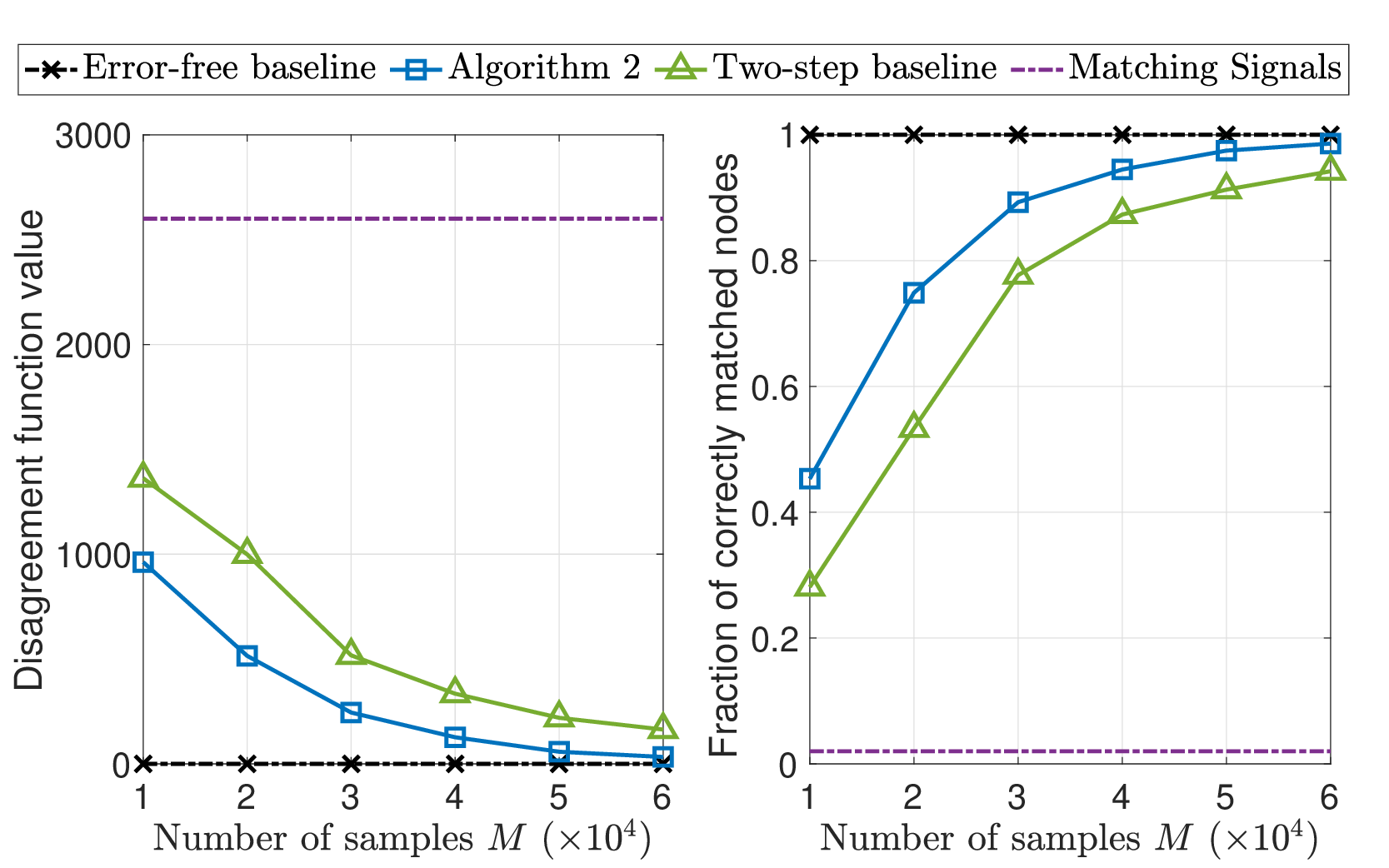}
				\caption{Matching of signals generated with the low-pass and high-pass filters.}
				\label{figa}
\end{figure}
   
Finally, we emphasize that the graph filters applied to different graphs generally differ and thus the resulting signals exhibit distinct covariance structures, rendering direct matching of graph signals impractical. In contrast, our method utilizes the eigenvectors of the signal covariance matrices, which unveils the spectral structure of the Laplacian matrices due to the intrinsic relationship between the graph filter and the graph Laplacian. To demonstrate this, we study a case where highly heterogeneous graph filters yield vastly different signal covariance matrices and different orders for their eigenvalues. Specifically, we match two ER graphs equipped with graph filters possessing distinct properties: Graph $\mathcal{G}_1$ utilizes a low-pass filter, $\mathcal{H}(\Lv^{(1)})=(\Iv_n+0.1\Lv^{(1)})^{-1}$, whereas $\mathcal{G}_2$ employs a high-pass polynomial filter, $\mathcal{H}(\Lv^{(2)})=\Lv^{(2)}+0.01(\Lv^{(2)})^2-4\Iv_n$. The other simulation parameters remain consistent with those in Fig. \ref{fig_M_ER}. 
 Fig. \ref{figa} compares the matching accuracy of our proposed method with the two-step baseline method. In addition, we include a baseline strategy of direct graph signal matching, which maximizes the similarity between the two sets of graph signals subject to a permutation matrix, $\text{tr}(\Pv\Yv_1\Yv_2^T)$, by using the Hungarian method. Here, $\Yv_i=[\yv_1^{(i)},\cdots,\yv_M^{(i)}],i=1,2,$ is the stack of the observed signals corresponding to each graph.
   Given the substantial difference between the two graph filters, direct graph signal matching proves impractical.
In contrast, our approach achieves precise graph matching with an adequate number of signal samples.

\subsection{Results on Real Networks}
\begin{figure}[!t]
	\centering
	\includegraphics[width=2.7 in]{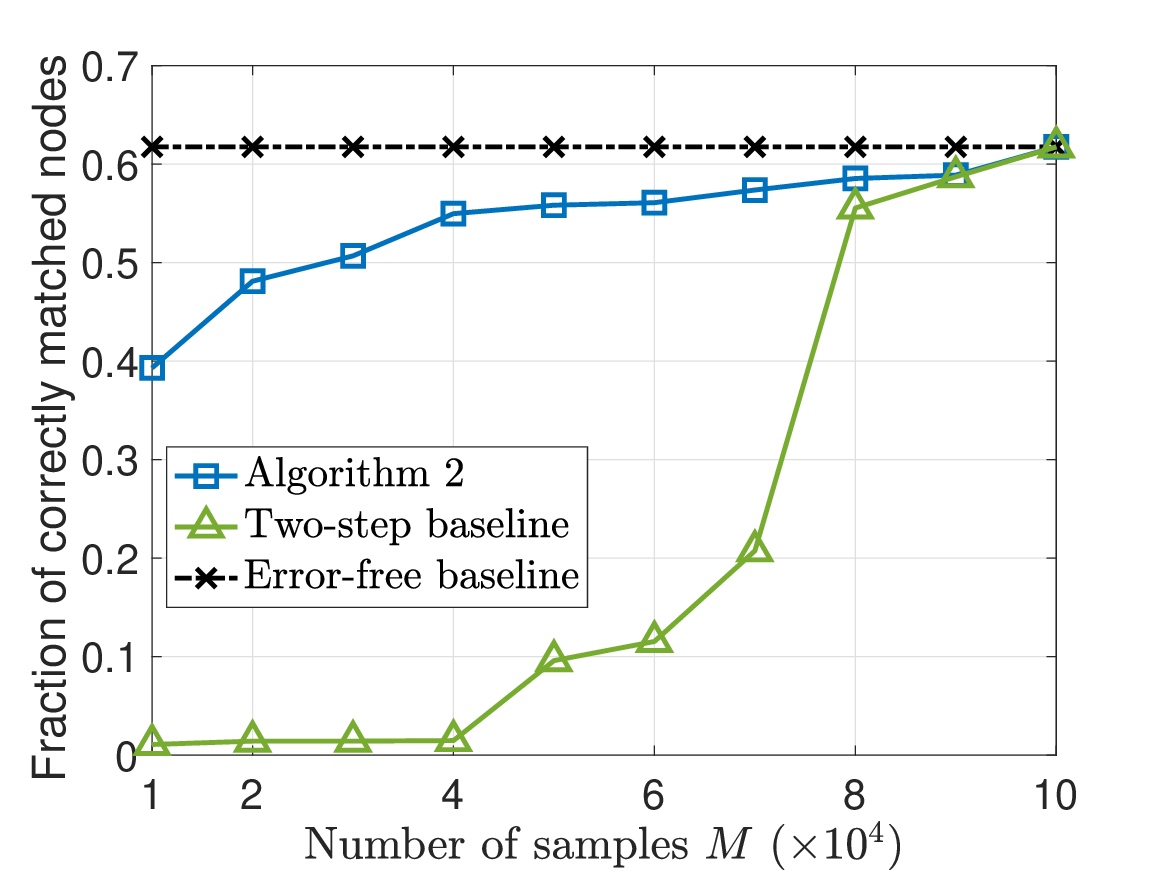}
	\caption{The fraction of correctly matched nodes on the \emph{HighSchool} network.}
	\label{fig_HighSchool}
\end{figure}
\begin{figure}[!t]
	\centering
	\includegraphics[width=2.7 in]{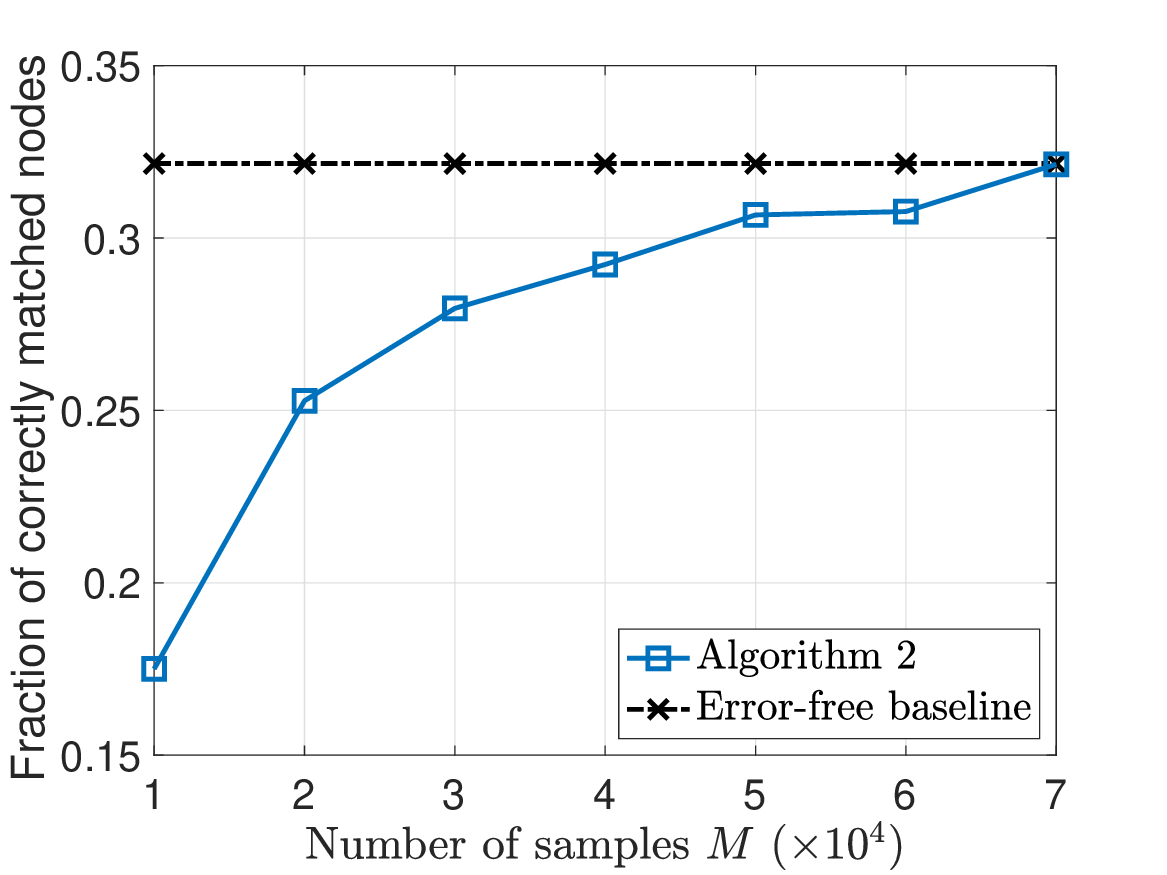}
	\caption{The fraction of correctly matched asymmetric nodes on the \emph{Facebook} network.}
	\label{fig_Facebook}
\end{figure}

We examine the blind matching of two subgraphs independently sampled from the \emph{HighSchool} network, as shown in Fig. \ref{fig_HighSchool}. The simulation parameters can be found in Section \ref{sec5_a}. 
The error-free baseline in \eqref{eq05} achieves a node-matching accuracy of approximately $62\%$. Meanwhile, our proposed blind method achieves comparable accuracy to this ideal baseline when $M\geq 10^5$. 

Moreover, we study blind graph matching of two \emph{symmetric} subgraphs sampled from the \emph{Facebook} network in Fig. \ref{fig_Facebook}. Since the computational cost of the graph inference solver increases significantly when $n$ exceeds several hundred, we only present the error-free baseline in Fig. \ref{fig_Facebook}. By examining the symmetry using the method in Remark \ref{remark1}, we find that $45$ out of $348$ nodes are symmetric subject to single swaps, making the matching problem not identifiable. Here, we apply our algorithm to the sampled graphs $\mathcal{G}_1$ and $\mathcal{G}_2$ with all the $348$ nodes and evaluate the faction of correct matching for the remaining $303$ nodes. While our algorithm and analysis are primarily designed for matching asymmetric graphs, we can identify over $30\%$ of the nodes over the symmetric graphs. 

We note that even with known graph topology, identifying all the symmetric nodes in a graph $\mathcal{G}$ is computationally expensive, as it involves finding all permutations $\Pv\in\mathcal{S}_n$ that satisfy $\text{dis}_{\mathcal{G}\to\mathcal{G}}(\Pv)=0$. For blind graph matching, the identification of symmetric structures with unknown graph topology becomes even more challenging. We envision that the analysis in Section \ref{sec4} provides a heuristic for approximately determining symmetric nodes of underlying graphs. Specifically, we expect that $c_j\geq \ell_j$ holds with high probability for any asymmetric node $j\in [n]$ in \eqref{eq23}--\eqref{eq23b} with a large $K$. In contrast, when node $j$ is symmetric, the value of $c_j$ is likely to be close to $\ell_j$. Inspired by this, we can approximately identify the symmetric nodes by estimating $c_j$   and  $\ell_j$ with the unavailable true eigenvectors $\Vv_k^{(i)}$ replaced by its estimate $\Uv_k^{(i)}$. However, we acknowledge that this problem requires further research.
\subsection{Comparisons With Shuffled Linear Regression}
					\begin{figure}[!t]
					\centering
					\includegraphics[width=3 in]{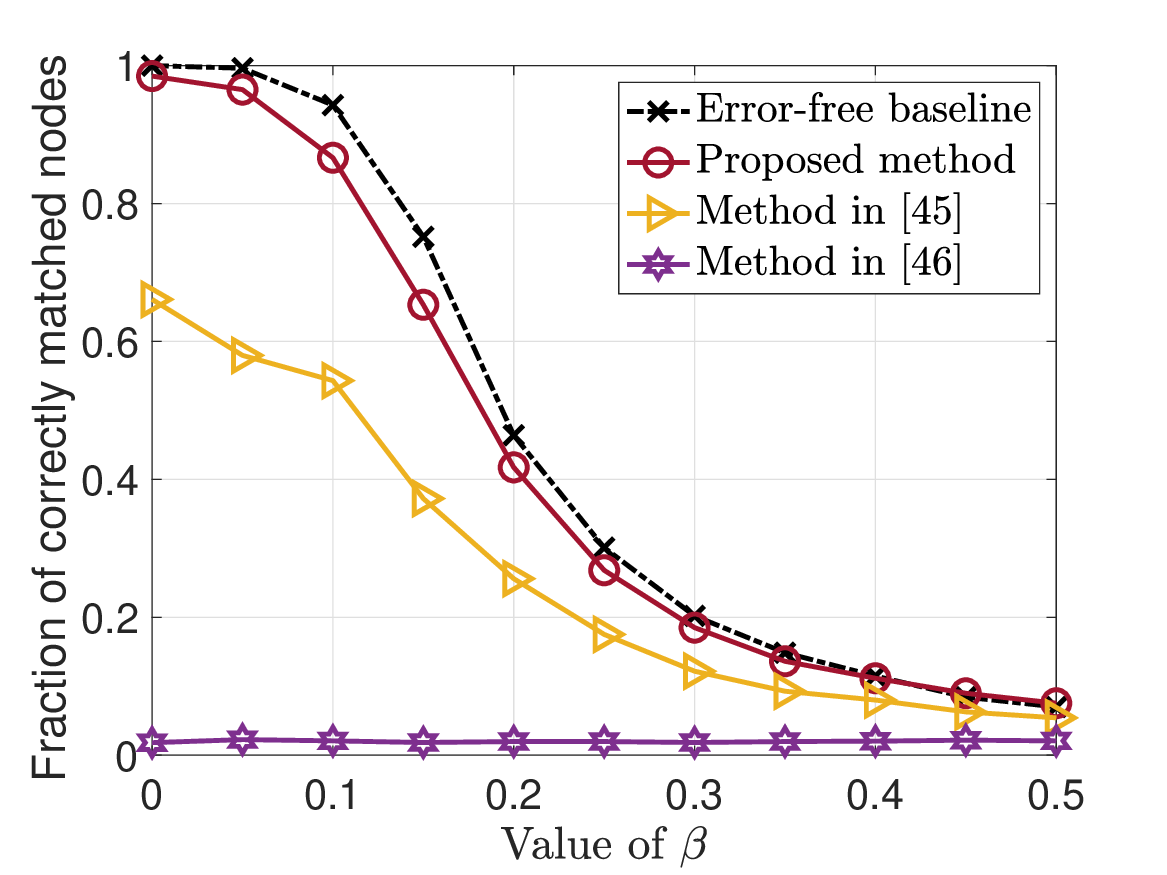}
					\caption{Comparison with the SLR algorithms in terms of the graph matching accuracy under the correlated Gaussian model.}
					\label{fig_slr}
				\end{figure}
The proposed graph matching algorithm can be viewed as an instance of the shuffled linear regression (SLR) problem \cite{SLR_ref1,SLR_ref2,SLR_ref3,SLR_ref4}. Specifically, by viewing the sample eigenvectors $\Uv_K^{(i)}$ in \eqref{eqeigen} as the noisy estimates of the true eigenvectors $\Vv_K^{(i)}$, we can represent them as
			\begin{align}\label{eq_r2}
					\Uv^{(1)}_K&=\Vv^{(1)}_K+\Nv_1,\nonumber\\
				\Uv^{(2)}_K&=(\Pv^\star)^T\Vv^{(1)}_K\Rv+\Nv_2,
				\end{align}
    where $\Pv^\star$ is the true permutation; $\Nv_1$ and $\Nv_2$ are error matrices; and $\Rv\in\Real^{K\times K}$ denotes an unknown transformation between the two noiseless eigenvectors defined as $\Vv^{(2)}_K\approx(\Pv^\star)^T\Vv^{(1)}_K\Rv$. Notably, when the two graphs are isomorphic, $\Rv$ is a diagonal matrix representing the sign ambiguity in the eigendecomposition. As a result, our solution in \eqref{eq12} can be viewed as a solver to the SLR problem of estimating $\Pv^\star$ from $\Uv^{(i)}_K$, which maximizes a similarity objective w.r.t. the absolute value of these sample eigenvectors.

    In this section, we numerically compare our method with the SLR algorithms in \cite{SLR_ref1,SLR_ref2}.  Specifically, \cite{SLR_ref1} focuses on alternatively denoising $\Vv_K^{(1)}$ and $\Rv$, followed by computing $\widehat \Pv$ by the linear assignment problem in \cite[Page 15]{SLR_ref1}. Meanwhile, the method in  \cite[Section 2.3]{SLR_ref2} estimates  $\widehat \Pv$ and $\Vv_K^{(1)}$ directly by sorting $\Uv_K^{(2)}$. 

  We simulate the Gaussian model in \eqref{eq28}. A larger $\b$ in \eqref{eq28} represents increasingly distinct underlying graphs, with $\b=0$ representing the isomorphic matching. Fig. \ref{fig_slr} plots the matching performance versus the value of $\b$. The simulation setup is consistent with that used for Fig. \ref{fig_b_GW}, and both the linear assignment problems in \eqref{eq12} and \cite{SLR_ref1} are solved using the Hungarian method \cite{Hungarian}. We see that our method outperforms the approach in \cite{SLR_ref1}, even at high values of $\beta$. Note that our method relies on the ordered sample eigenvectors of the graph filters for node matching. Notably, even when the two graph filters significantly differ, accurate eigenvector estimation can be achieved with a sufficient number of signal samples. On the other hand, the method in \cite{SLR_ref2} struggles to achieve precise node matching as it does not leverage the information from $\Uv^{(1)}_K$. 

\section{Conclusions}\label{sec6}
In this work, we studied blind graph matching using graph signals for two graphs with unknown topologies. 
We extended the conventional spectral method by using the selected eigenbases of the sample covariance matrices. Our method relies on the assumption that the two graph filters have the same characteristic and preserve the same order of filter responses. We theoretically analyzed the error in blind matching and proved that our method achieves accurate graph matching with sufficiently many signal samples and small signal noise. Numerical results on synthetic data and real networks confirm the efficiency of the proposed algorithm.

Our work demonstrates the effectiveness of directly matching graphs using graph signals, opening up two interesting directions for future research. First, it is interesting to investigate blind graph matching with generally unknown filter characteristics. Second, our work highlights the rich information that graph signals can provide about the underlying graph structure. It is worthwhile to incorporate privacy-preserving mechanisms into blind graph matching systems to protect the private information of individual nodes.
\appendices
\section{Proof of Proposition  \ref{pro2}}
\label{appb}

Define $\Ev\triangleq \overline\Uv_K^{(1)}(\overline\Uv_K^{(2)})^T-\overline\Vv_K^{(1)}(\overline\Vv_K^{(2)})^T$ as the perturbation matrix of the cost matrix in \eqref{eq12}. We have
\begin{align}\label{appa01}
	&~~~~K-\tr(	\widehat\Pv^T\overline\Vv_K^{(1)}(\overline\Vv_K^{(2)})^T)\nonumber\\
	&	= K-\tr(	\widehat\Pv^T\overline\Uv_K^{(1)}(\overline\Uv_K^{(2)})^T)+\tr(\widehat\Pv^T\Ev)\nonumber\\
	&	\overset{(a)}{\leq} K-\tr(	(\Pv^\star)^T\overline\Uv_K^{(1)}(\overline\Uv_K^{(2)})^T)+\tr(\widehat\Pv^T\Ev)\nonumber\\
	%&=\sum_{i=1}^n \left( (\overline\nuv^{(1)}_i)^T\overline\nuv_{\pi(i)}^{(2)}-(\overline\muv^{(1)}_i)^T\overline\muv_{\pi(i)}^{(2)}\right) +\tr(\widehat\Pv^T\Ev)\nonumber\\
	%	&=K-\sum_{j=1}^n\sum_{k=1}^K |u_{jk}^{(1)}||u_{\pi^\star(j)k}^{(2)}|+\tr(\widehat\Pv^T\Ev)\nonumber\\
	&=K-\sum_{k=1}^K (\overline\uv^{(1)}_k)^T\Pv^\star\overline\uv_{k}^{(2)} +\tr(\widehat\Pv^T\Ev)\nonumber\\
	&\overset{(b)}{\leq} K-\sum_{k=1}^K \left|(\uv^{(1)}_k)^T\Pv^\star\uv_{k}^{(2)}\right|+n\norm{\Ev}_{\max},
\end{align}
where 
%$\pi^\star(\cdot)$ is the node mapping function associated with $\Pv^\star$, 
$(a)$ is because $\widehat \Pv$ maximizes \eqref{eq12}, $(b)$ follows from the triangle inequality, and $\norm{\Ev}_{\max}=\max_{ij}|[\Ev]_{ij}|$ is the max norm of $\Ev$.

%Eq. \eqref{eq21} shows that the `optimality gap' is upper bounded by two terms: The first term $k-\sum_{k=1}^K \left|(\uv^{(1)}_k)^T\Pv^\star\uv_{k}^{(2)}\right|$ measures the mismatch in the two sample eigenvectors, and the second term $n\norm{\Ev}_{\max}$ bounds the maximum perturbation between the cost matrices of \eqref{eq12} and \eqref{eq05}. 
For any two vectors $\xv$ and $\yv$ of the same dimension, we denote their angle by $\angle(\xv,\yv)\triangleq \arccos(\frac{\xv^T\yv}{\norm{\xv}_2\norm{\yv}_2})$. To control the two error terms in \eqref{appa01}, the next two results follow. 

\lemma{\label{lemma3}Under the conditions of Proposition \ref{pro2}, we have
	\begin{align}\label{appa02}
		&K-\sum_{k=1}^K \left|(\uv^{(1)}_k)^T\Pv^\star\uv_{k}^{(2)}\right|\nonumber\\
		\leq& \sum_{k=1}^K\left( \sin(\angle(\uv^{(1)}_k,\vv^{(1)}_k))+\sin(\angle(\uv^{(2)}_k,\vv^{(2)}_k))\right)^2.
	\end{align}
}
\begin{IEEEproof}
	See Appendix \ref{appd}.
\end{IEEEproof}

\lemma{\label{lemma4}Under the conditions of Proposition \ref{pro2}, we have \eqref{appa03} shown on top of the next page.
	\begin{figure*}[ht]
		\begin{align}\label{appa03}
			&	\norm{\Ev}_{\max}\leq 2\sqrt{	\sum_{k=1}^K \sin^2(\angle(\uv^{(1)}_k,\vv^{(1)}_k))}\sqrt{	\sum_{k=1}^K \sin^2(\angle(\uv^{(2)}_k,\vv^{(2)}_k))}
			%		\sum_{k=1}^K\left( \sin^2(\angle(\uv^{(1)}_k,\vv^{(1)}_k))+\sin^2(\angle(\uv^{(2)}_k,\vv^{(2)}_k))\right)\nonumber\\
			+\sqrt{2}\left( \sqrt{	\sum_{k=1}^K \sin^2(\angle(\uv^{(1)}_k,\vv^{(1)}_k))}+\sqrt{	\sum_{k=1}^K \sin^2(\angle(\uv^{(2)}_k,\vv^{(2)}_k))}\right).
		\end{align}
		\hrulefill
	\end{figure*}
	
}
\begin{IEEEproof}
	See Appendix \ref{appe}.
\end{IEEEproof}
Applying the variant of the Davis-Kahan theorem in \cite[Corollary 3]{DK_theorem2}, for any $k\in [K]$, we have
\begin{align}\label{temp3}
	\sin(\angle(\uv^{(i)}_k,\vv^{(i)}_k))\leq \frac{2\norm{\Deltav^{(i)}}_2}{\delta_k^{(i)}}\leq \frac{2\norm{\Deltav^{(i)}}_2}{\delta_{\min,K}},
\end{align}
where $\Delta^{(i)}$ and $ \delta_{\min,K}$ are defined in Proposition \ref{pro2}.
Combining \eqref{appa01}--\eqref{temp3}, we have \eqref{eq21}.

\section{Proof of Proposition \ref{pro3}}
\label{appc}
When $\widehat \Pv\neq \Pv^\star$, define $\mathcal{T}\triangleq\{j\in[n]:\hat \pi(j)\neq \pi^\star(j)\}$. 
%It can be verified that there exists a bijective node mapping $\tilde\pi(\cdot):\mathcal{T}\to\mathcal{T}$ such that $\hat \pi(j)= \tilde\pi(\pi^\star(j)),\forall j\in \mathcal{T}$. 
Denoting $\Xv\triangleq \overline\Uv^{(1)}_K(\overline\Uv^{(2)}_K)^T$, we have
\begin{align}\label{appc01}
	\widehat \Pv\neq \Pv^\star
	\Rightarrow&\widehat \Pv^T\overline\Uv^{(1)}_K(\overline\Uv^{(2)}_K)^T\geq (\Pv^\star)^T\overline\Uv^{(1)}_K(\overline\Uv^{(2)}_K)^T\nonumber\\
	\Leftrightarrow&\sum_{j\in\mathcal{T} }x_{\hat \pi(j),j}-x_{ \pi^\star(j),j}\geq0
\end{align}
Recall that $\Ev\triangleq \overline\Uv_K^{(1)}(\overline\Uv_K^{(2)})^T-\overline\Vv_K^{(1)}(\overline\Vv_K^{(2)})^T$. We have
\begin{align}\label{appc02}
	&x_{ \pi^\star(j),j}=\left[\overline\Vv_K^{(1)}(\overline\Vv_K^{(2)})^T\right]_{ \pi^\star(j),j}+e_{ \pi^\star(j),j}\geq c_j-\norm{\Ev}_{\max},\\
	&x_{\hat \pi(j),j}\overset{(a)}{\leq} c_j-\rho+e_{ \pi^\star(j),j}\leq c_j-\rho+\norm{\Ev}_{\max},\label{appc03}
\end{align}
where $\norm{\Ev}_{\max}=\max_{ij}|[\Ev]_{ij}|$ is the max norm of $\Ev$, and $(a)$ is from the definition of $\rho$ in \eqref{eq24}. Plugging \eqref{appc02} and \eqref{appc03} into \eqref{appc01}, we have
\begin{align}
	\eqref{appc01}\Rightarrow |\mathcal{T}|\left( 2\norm{\Ev}_{\max}-\rho\right) \geq0.
\end{align}
Therefore, the error probability is bounded by
\begin{align}\label{appc08}
	\Pr(\widehat \Pv\neq \Pv^\star)\leq \Pr\left( 	\norm{\Ev}_{\max}\geq\frac{\rho}{2}\right). 
\end{align}
Applying the results in Lemma \ref{lemma4}, \eqref{temp3} and Lemma \ref{lemma5}, for sufficiently large $M$, we have
\begin{align}\label{appc07}
	&~~~~\norm{\Ev}_{\max}\nonumber\\
	&\leq	\frac{2\sqrt{2K}}{\delta_{\min,K}}( \norm{\Deltav^{(1)}}_2+\norm{\Deltav^{(2)}}_2+\frac{2\sqrt{2K}}{\delta_{\min,K}}\norm{\Deltav^{(1)}}_2\norm{\Deltav^{(2)}}_2)\nonumber\\
	&\leq \frac{2\sqrt{2K}}{\delta_{\min,K}}( \norm{\Deltav^{(1)}}_2+\norm{\Deltav^{(2)}}_2)+\frac{{4K}}{\delta^2_{\min}(K)}(\norm{\Deltav^{(1)}}_2^2+\norm{\Deltav^{(2)}}^2_2)\nonumber\\
	&\overset{(a)}{\leq}\frac{2\sqrt{2K}\delta_{\min,K}+4K}{\delta^2_{\min}(K)}( \norm{\Deltav^{(1)}}_2+\norm{\Deltav^{(2)}}_2),
\end{align}
where $(a)$ follows from $\norm{\Deltav^{(i)}}_2\leq 1$ for sufficiently large $M$ (cf. Lemma \ref{lemma5}). Substituting \eqref{appc07} into \eqref{appc08}, we have
\begin{align}\label{appc04}
	%	&~~~~\Pr(\widehat \Pv\neq \Pv^\star)\nonumber\\
	&\eqref{appc08}\leq \Pr\left( \norm{\Deltav^{(1)}}_2+\norm{\Deltav^{(2)}}_2\geq \frac{\rho \delta^2_{\min}(K)}{8K+4\sqrt{2K}\delta_{\min,K}}\right),
\end{align}

To  further bound \eqref{appc04}, the next lemma follows.
\lemma{\label{lemma6}Let $x$ and $y$ be two random variables and $t$ be any real number. For any $\zeta\in[0,1]$,
	\begin{align}
		&	\Pr(x+y\geq t)\leq \Pr(x\geq \zeta t)+\Pr(y\geq (1-\zeta)t),
	\end{align}
}
\begin{IEEEproof}Applying the law of total probability, we have
	\begin{align}
		&\Pr(x+y\geq t)\nonumber\\
		&=	\Pr(x+y\geq t|y\geq (1-\zeta)t)\Pr(y\geq (1-\zeta)t)\nonumber\\
		&~~+	\Pr(x+y\geq t|y< (1-\zeta)t)\Pr(y< (1-\zeta)t)\nonumber\\
		&\leq \Pr(y\geq (1-\zeta)t)+	\Pr(x+y\geq t,y< (1-\zeta)t)\nonumber\\
		&\leq \Pr(y\geq (1-\zeta)t)+	\Pr(x\geq \zeta t).
	\end{align}
	
\end{IEEEproof}	

Let  $\omega\triangleq  \frac{\rho \delta^2_{\min}(K)}{8K+4\sqrt{2K}\delta_{\min,K}}$ and applying Lemma \ref{lemma6}, we have 
\begin{align}\label{eqappa1}
	&\Pr(\widehat \Pv\neq \Pv^\star)\nonumber\\
	\leq &\min_{\zeta\in[0,1]} \left( \Pr(\norm{\Deltav^{(1)}}_2\geq\zeta\omega)+\Pr(	\norm{\Deltav^{(2)}}_2\geq(1-\zeta)\omega)\right) .
\end{align}

Applying Lemma \ref{lemma5}, for sufficiently large $M$ and any $t>0$, we have
$$\Pr\left(\norm{\Deltav^{(i)}}_2\geq \sigma^2+\sqrt{\frac{Cn\ln(n/t)}{M}}\right)\leq 1-2t,$$
where $C=\max_{i}C_i^2Y^2$. Let $t^\prime=\sigma^2+\sqrt{\frac{Cn\ln(n/t)}{M}}$. We conclude that for any $t^\prime>\sigma^2$,
\begin{align}\label{appc05}
	\Pr\left(\norm{\Deltav^{(i)}}_2\geq t^\prime\right) \leq 2ne^{-\frac{M(t^\prime-\sigma^2)^2}{nC}}.
\end{align}
For any $\frac{\sigma^2}{\omega}<\zeta<1-\frac{\sigma^2}{\omega}$, substituting $t^\prime=\zeta\omega$ and  $t^\prime=(1-\zeta)\omega$ into \eqref{appc05}, we have
\begin{align}\label{appc06}
	\eqref{eqappa1}\leq& 2n\min_{\frac{\sigma^2}{\omega}<\zeta<1-\frac{\sigma^2}{\omega}}\underbrace{ e^{-\frac{M(\zeta\omega-\sigma^2)^2}{nC}}+e^{-\frac{M((1-\zeta)\omega-\sigma^2)^2}{nC}}}_{\triangleq g(\zeta)}.
\end{align}
Note that $g(\frac{\sigma^2}{\omega})=g(1-\frac{\sigma^2}{\omega})=1+e^{-\frac{M(\omega-2\sigma^2)^2}{nC}}$. Moreover, the derivative of $g(\zeta)$ is given by
\begin{align}
	g^\prime(\zeta)=&\frac{2M\omega}{nC}\big(((1-\zeta)\omega-\sigma^2)e^{-\frac{M((1-\zeta)\omega-\sigma^2)^2}{nC}}\nonumber\\
	& -(\zeta\omega-\sigma^2)e^{-\frac{M(\zeta\omega-\sigma^2)^2}{nC}}\big).
\end{align}
For sufficiently large $M$, the function of $xe^{-\frac{M}{nC}x^2}$ is decreasing with its argument $x$. Therefore, we have $	g^\prime(\zeta)<0$ when $\zeta<\half$ and $	g^\prime(\zeta)>0$ when $\zeta>\half$, implying that $\min_\zeta g(\zeta)=g(\half)=2e^{-\frac{M(\omega/2-\sigma^2)^2}{nC}}$. Combining this result with \eqref{appc06} completes the proof.

\section{Proof of Lemma \ref{lemma3}}
\label{appd}

Fixing the eigendecomposition in \eqref{eq11}, there exist eigendecompositions in  \eqref{eqeigen} such that 
%we can choose 
the signs of the eigenvectors 
%$\vv_k^{(1)}$ and $\vv_{k}^{(2)}$ such that  $(\Vv_K^{(1)})^T\vv_{\pi^\star(k)}^{(2)}\geq 0$ without loss of generality. Similarly, we choose the signs of 
%
$\uv^{(1)}_k$ and $\uv_{k}^{(2)}$ satisfy
$(\uv^{(1)}_k)^T\vv^{(1)}_k\geq 0$ and $(\uv_{k}^{(2)})^T\vv_{k}^{(2)}\geq 0$ for $\forall k$.
%As a result, we have $\Vv_K^{(1)}=\vv_{\pi^\star(k)}^{(2)}$ due to the graph isomorphism. 
When $\mathcal{G}_1$ and $\mathcal{G}_2$ are isomorphic, it follows that the eigenvectors of their Laplacian matrices are identical subject to the true permutation matrix $\Pv^\star$, i.e., $\vv_{k}^{(1)}=\Pv^\star\vv_{k}^{(2)},\forall k$.
Consequently, we have 
\begin{align}\label{temp1}
	&	\left|(\uv^{(1)}_k)^T\Pv^\star\uv_{k}^{(2)}\right|\nonumber\\
 =&\left|(\uv^{(1)}_k+\vv^{(1)}_k-\vv^{(1)}_k)^T\Pv^\star(\uv_{k}^{(2)}-\vv_{k}^{(2)}+\vv_{k}^{(2)})\right|\nonumber\\
	=	&	\big|(\uv^{(1)}_k)^T\vv^{(1)}_k+(\vv_{k}^{(2)})^T\uv_{k}^{(2)}-1\nonumber\\
	&~~~~+(\uv^{(1)}_k-\vv^{(1)}_k)^T(\Pv^\star\uv_{k}^{(2)}-\Pv^\star\vv_{k}^{(2)})\big|\nonumber\\
	\overset{(a)}{\geq} &(\uv^{(1)}_k)^T\vv^{(1)}_k+(\uv_{k}^{(2)})^T\vv_{k}^{(2)}-1\nonumber\\
	&~~~~-\big|(\uv^{(1)}_k-\vv^{(1)}_k)^T(\Pv^\star\uv_{k}^{(2)}-\Pv^\star\vv_{k}^{(2)})\big|\nonumber\\
	\overset{(b)}{\geq} &(\uv^{(1)}_k)^T\vv^{(1)}_k+(\uv_{k}^{(2)})^T\vv_{k}^{(2)}-1\nonumber\\
	&~~~~-\norm{\uv^{(1)}_k-\vv^{(1)}_k}_2\norm{\uv_{k}^{(2)}-\vv_{k}^{(2)}}_2,
\end{align}
where $(a)$ follows from the triangle inequality and $(b)$ follows from the 
Cauchy–Schwarz inequality. 
Substituting \eqref{temp1} into \eqref{appa01} and applying the definition of the vector angle, we have
\begin{align}\label{temp2}
	&K-\sum_{k=1}^K \left|(\uv^{(1)}_k)^T\Pv^\star\uv_{k}^{(2)}\right|\nonumber\\
	\leq&2K-\sum_{k=1}^K\left( \cos(\angle(\uv^{(1)}_k,\vv^{(1)}_k))+\cos(\angle(\uv^{(2)}_k,\vv^{(2)}_k))\right) \nonumber\\
	&~~~~+\sum_{k=1}^K\norm{\uv^{(1)}_k-\vv^{(1)}_k}_2\norm{\uv_{k}^{(2)}-\vv_{k}^{(2)}}_2\nonumber\\
	\overset{(a)}{\leq} &2K-\sum_{k=1}^K\left( \cos^2(\angle(\uv^{(1)}_k,\vv^{(1)}_k))+\cos^2(\angle(\uv^{(2)}_k,\vv^{(2)}_k))\right) \nonumber\\
	&~~~~+2\sum_{k=1}^K\sin(\angle(\uv^{(1)}_k,\vv^{(1)}_k))\sin(\angle(\uv^{(2)}_k,\vv^{(2)}_k))\nonumber\\
	=&\sum_{k=1}^K\left( \sin(\angle(\uv^{(1)}_k,\vv^{(1)}_k))+\sin(\angle(\uv^{(2)}_k,\vv^{(2)}_k))\right)^2,
	%	\nonumber\\
	%		\overset{(b)}{\leq} &2\sum_{k=1}^K\left( \sin^2(\angle(\uv^{(1)}_k,\vv^{(1)}_k))+\sin^2(\angle(\uv^{(2)}_k,\vv^{(2)}_k))\right),
\end{align}
where $(a)$ is because $\cos(\angle(\uv^{(i)}_k,\vv^{(i)}_k))\leq 1$ and $\norm{\uv^{(i)}_k-\vv^{(i)}_k}_2\leq \sqrt{2}\sin(\angle(\uv^{(i)}_k,\vv^{(i)}_k))$.
% and $(b)$ is from the inequality of arithmetic and geometric means.

\section{Proof of Lemma \ref{lemma4}}
\label{appe}
Let $\norm{\Ev}_{\max}=|[\Ev]_{j^\star l^\star}|$ for some $(j^\star,l^\star)=\argmax_{(j,l)}|[\Ev]_{jl}|$. Then, we have
\begin{align}\label{pa23}
	\norm{\Ev}_{\max}=|E_{j^\star l^\star}|=\left|\sum_{k=1}^K \left( |u_{j^\star k}^{(1)}u_{l^\star k}^{(2)}|-|v_{j^\star k}^{(1)}v_{l^\star k}^{(2)}|\right)\right|.
\end{align}
The eigendecompositions of $\widehat\Cv_y^{(i)}$ and $\Cv_y^{(i)}$ in \eqref{eq11} and \eqref{eqeigen} inherently exhibit sign ambiguities. Consequently, there are eigendecompositions for which $u_{j^\star k}^{(1)}u_{l^\star k}^{(2)}\geq 0$ and $v_{j^\star k}^{(1)}v_{l^\star k}^{(2)}\geq 0$ hold for $\forall k$. Define $\delta\Vv_K^{(i)}\triangleq \Uv_K^{(i)}-\Vv_K^{(i)}$. We have
\begin{align}\label{appe1}
	&	\norm{\Ev}_{\max}= \left|\sum_{k=1}^K \left( u_{j^\star k}^{(1)}u_{l^\star k}^{(2)}-v_{j^\star k}^{(1)}v_{l^\star k}^{(2)}\right)\right|\nonumber\\
	&=\left|[\Uv_K^{(1)}(\Uv_K^{(2)})^T-\Vv_K^{(1)}(\Vv_K^{(2)})^T]_{j^\star l^\star}\right|\nonumber\\
	&\leq \norm{\Uv_K^{(1)}(\Uv_K^{(2)})^T-\Vv_K^{(1)}(\Vv_K^{(2)})^T}_{\max}\nonumber\\
	&\leq \norm{\Uv_K^{(1)}(\Uv_K^{(2)})^T-\Vv_K^{(1)}(\Vv_K^{(2)})^T}_{2}\nonumber\\
%	&=\norm{\delta\Vv_K^{(1)}(\Vv_K^{(2)})^T+\Vv_K^{(1)}(\delta\Vv_K^{(2)})^T+\delta\Vv_K^{(1)}(\delta\Vv_K^{(2)})^T}_{2}\nonumber\\
	&\leq \norm{\delta\Vv_K^{(1)}}_{2}+\norm{\delta\Vv_K^{(2)}}_{2}+\norm{\delta\Vv_K^{(1)}}_2\norm{\delta\Vv_K^{(2)}}_{2}	\nonumber\\
	&\leq \norm{\delta\Vv_K^{(1)}}_{F}+\norm{\delta\Vv_K^{(2)}}_{F}+\norm{\delta\Vv_K^{(1)}}_F\norm{\delta\Vv_K^{(2)}}_{F}.
	%	&\leq \norm{\delta\Vv_K^{(1)}}_{F}+\norm{\delta\Vv_K^{(2)}}_{F}+\half(\norm{\delta\Vv_K^{(1)}}_F^2+\norm{\delta\Vv_K^{(2)}}_{F}^2).
\end{align}
Note that
\begin{align}\label{appe2}
	\norm{\delta\Vv_K^{(1)}}_{F}^2=
%	&2K-2\tr\left((\Uv_K^{(i)})^T\Vv_K^{(i)}\right)\nonumber\\
	&2K-2\sum_{k=1}^K \cos(\angle(\uv^{(i)}_k,\vv^{(i)}_k))\nonumber\\
	\leq &2K-2\sum_{k=1}^K \cos^2(\angle(\uv^{(i)}_k,\vv^{(i)}_k))\nonumber\\
	=&2\sum_{k=1}^K \sin^2(\angle(\uv^{(i)}_k,\vv^{(i)}_k)).
\end{align}
Combining \eqref{appe1} and \eqref{appe2} completes the proof.

%\begin{footnotesize}
\bibliographystyle{IEEEtran}
\bibliography{IEEEabrv,ref}
%\end{footnotesize}
%

\end{document}

%% file: begin.tex
\newif\ifhavebib

\input{Chead.tex}

\input{Chead2.tex}
\input{defns.tex}

%% file: Chead.tex
\usepackage{blindtext}
\usepackage{graphicx}
\usepackage{graphicx}
\usepackage{array}
\usepackage{url}
\usepackage{algorithmic}
\usepackage[cmex10]{amsmath}
\usepackage{ifpdf}
\usepackage{amssymb,amsmath}
\usepackage{empheq}
\usepackage{stfloats}  
\usepackage{textcomp}
\usepackage{cite}
\usepackage{multirow}
\usepackage{diagbox}
\usepackage{subcaption}
\usepackage{mathrsfs}
\let\oldFootnote\footnote
\newcommand\nextToken\relax

\renewcommand\footnote[1]{%
    \oldFootnote{#1}\futurelet\nextToken\isFootnote}

\newcommand\isFootnote{%
    \ifx\footnote\nextToken\textsuperscript{,}\fi}

\usepackage[ruled,vlined,lined,ruled,boxed]{algorithm2e}
\usepackage[dvipsnames,usenames]{color}
\usepackage[normalem]{ulem}
\usepackage{amsthm}

\usepackage{graphicx}
\usepackage{ifpdf}
\usepackage{cite}
\usepackage{enumerate}
\usepackage{enumitem}

\usepackage{array}
\usepackage{mdwmath}
\usepackage{mdwtab}
\usepackage{eqparbox}
\usepackage{bm}

\usepackage{setspace}
\usepackage{booktabs}
\usepackage[subnum]{cases}
\usepackage{rotating}

\usepackage{listings} 
\definecolor{Red}{rgb}{1,0,0}
\definecolor{Blue}{rgb}{0,0,1}
\definecolor{Olive}{rgb}{0.41,0.55,0.13}
\definecolor{Green}{rgb}{0,1,0}
\definecolor{MGreen}{rgb}{0,0.8,0}
\definecolor{DGreen}{rgb}{0,0.55,0}
\definecolor{Yellow}{rgb}{1,1,0}
\definecolor{Cyan}{rgb}{0,1,1}
\definecolor{Magenta}{rgb}{1,0,1}
\definecolor{Orange}{rgb}{1,.5,0}
\definecolor{Violet}{rgb}{.5,0,.5}
\definecolor{Purple}{rgb}{.75,0,.25}
\definecolor{Brown}{rgb}{.75,.5,.25}
\definecolor{Grey}{rgb}{.5,.5,.5}

\newcommand{\boxhead}[5]{
   \pagestyle{myheadings}
   \thispagestyle{plain}
   \setcounter{page}{1}
   \noindent
   \begin{center}
   \framebox{
      \vbox{\vspace{2mm}
    \hbox to 6.28in { {\bf #1 \hfill} }
       \vspace{6mm}
       \hbox to 6.28in { {\Large \hfill \bf #2  \hfill} }
       \vspace{6mm}
       \hbox to 6.28in { {\it #3 #4 \hfill  #5} }
      \vspace{2mm}}
   }
   \end{center}
   \markboth{#5 -- #2}{#5 -- #2}
   \vspace*{4mm}
}

%% file: Chead2.tex
\usepackage[colorlinks=true, urlcolor=black,  linkcolor=black, citecolor=black]{hyperref}
\theoremstyle{definition}
 
\newtheorem{proposition}{Proposition}

\newtheorem{lemma}{Lemma}

\newtheorem{assumption}{Assumption}
\newtheorem{remark}{Remark}

\newtheorem{example}{Example}

%% file: defns.tex
\DeclarePairedDelimiterX{\infdivx}[2]{(}{)}{%
	#1\;\delimsize\|\;#2%
}

\DeclarePairedDelimiter{\norm}{\lVert}{\rVert}
\DeclarePairedDelimiter{\abs}{\lvert}{\rvert}
\DeclareMathOperator*{\argmax}{\mathop{\arg\max}}
\DeclareMathOperator*{\argmin}{\mathop{\arg\min}}
\def\tr{\mathop{\rm tr}\nolimits}%
\def\diag{\mathop{\rm diag}\nolimits}%
\def\rank{\mathop{\rm rank}\nolimits}%
\renewcommand{\Pr}{\mathscr{P}}

\newcommand{\Cv}{{\bf C}}

\newcommand{\Xv}{{\bf X}}
\newcommand{\Yv}{{\bf Y}}
\newcommand{\Zv}{{\bf Z}}
\newcommand{\Uv}{{\bf U}}

\newcommand{\Vv}{{\bf V}}
\newcommand{\Rv}{{\bf R}}

\newcommand{\Av}{{\bf A}}

\newcommand{\Ev}{{\bf E}}
\newcommand{\Pv}{{\bf P}}

\newcommand{\Lv}{{\bf L}}
\newcommand{\Gv}{{\bf G}}
\newcommand{\Sv}{{\bf S}}
\newcommand{\Nv}{{\bf N}}
\newcommand{\Iv}{{\bf I}}

\newcommand{\xv}{{\bf x}}

\newcommand{\yv}{{\bf y}}

\newcommand{\uv}{{\bf u}}
\newcommand{\vv}{{\bf v}}

\newcommand{\wv}{{\bf w}}

\newcommand{\Gammav}{\boldsymbol \Gamma}

\newcommand{\muv}{\boldsymbol \mu}

\newcommand{\Lambdav}{\boldsymbol \Lambda}
\newcommand{\Deltav}{\boldsymbol \Delta}

\def\a{\alpha}
\def\b{\beta}

\DeclareMathOperator\E{E}

 \def\E{\mathbb{E}}
 
 \def\Pr{\mathrm{Pr}}
\def\de \mathrm{d}
\newcommand{\Norm}{\mathcal{N}}

\newcommand\eg{e.g.,\xspace}
\newcommand\ie{i.e.,\xspace}
\def\textiid{i.i.d.\@\xspace}

\newcommand\iid{\ifmmode\text{ i.i.d. } \else \textiid \fi}

\newcommand{\Real}{\mathbb{R}}
\newcommand{\half}{\frac{1}{2}}

\newcommand{\beqs}{\begin{equation*}}
\newcommand{\eeqs}{\end{equation*}}
\newcommand{\beq}{\begin{equation}}
\newcommand{\eeq}{\end{equation}}